\documentclass[letterpaper, 10 pt, conference]{ieeeconf}  

\IEEEoverridecommandlockouts                              
\usepackage{cite}
\usepackage{amsmath,amssymb,amsfonts}
\usepackage{algorithmic}
\usepackage{subcaption}
\usepackage{graphicx}
\usepackage{algorithm,algorithmic}
\usepackage{mathtools}
\usepackage{commath}
\usepackage{enumerate}

\newtheorem{definition}{Definition}
\newtheorem{lemma}{Lemma}
\newtheorem{remark}{Remark}

 \newtheorem{theorem}{Theorem}[section]

\newtheorem{assumption}{Assumption}
\newtheorem{problem}{Problem}

\usepackage{hyperref}
\hypersetup{hidelinks=true}

\title{\LARGE \bf
Online Model Order Reduction of Linear Systems via $(\gamma,\delta)$-Similarity
}

\author{Shivam Bajaj, Carolyn L. Beck and Vijay Gupta
\thanks{Shivam Bajaj and Vijay Gupta are with the Elmore Family School of Electrical and Computer Engineering, Purdue University, West Lafayette, USA (e-mail: bajaj41@purdue.edu, gupta869@purdue.edu). Carolyn L. Beck is with the Department of Industrial and Enterprise Systems Engineering, University of Illinois, Urbana-Champaign, Champaign, USA (e-mail: beck3@illinois.edu)}}%

\begin{document}

\maketitle
\thispagestyle{empty}
\pagestyle{empty}

\begin{abstract}

Model order reduction aims to determine a low-order approximation of high-order models with least possible approximation errors. 
    For application to physical systems, it is crucial that the reduced order model (ROM) is robust to any disturbance that acts on the full order model (FOM) -- in the sense that the output of the ROM remains a good approximation of 
    that of the FOM, even in the presence of such disturbances. 
    In this work, we present a framework for online model order reduction  for a class of continuous-time linear systems that ensures this property 
    for any $\mathcal{L}_2$ disturbance. 
    Apart from robustness to disturbances in this sense, the proposed framework also displays other desirable properties for model order reduction: (1) a provable bound on the error defined as the $\mathcal{L}_2$ norm of the difference between the output of the ROM and FOM, (2) preservation of stability, (3) compositionality properties and a provable error bound for arbitrary interconnected systems, (4) a provable bound on the output of the FOM when the controller designed for the ROM is used with the FOM, and finally, (5) compatibility with existing approaches such as balanced truncation and moment matching. Property (4) does not require computation of any gap metric and property (5) is beneficial as existing approaches can also be equipped with some of the preceding properties. 
    The theoretical results are corroborated on numerical case studies, including on a building model.
\end{abstract}

\section{Introduction}\label{sec:introduction}
Complex interconnected large-scale systems are ubiquitous; transportation networks, supply chain and logistics systems, water networks, connected autonomous vehicles, microelectromechanical systems, and biological systems such as the human brain or enzyme-substrate reactions are all examples of large-scale systems \cite{kordestani2021recent,briggs1925note}. Designing accurate models of such systems is crucial but extremely difficult as such systems often have very high number of dimensions or degrees of freedom. Model order reduction is a widely studied approach for dealing with the issue of model complexity, with balanced truncation and moment matching being two of the most popular methods. 

While the primary aim of model order reduction is to determine a lower-dimensional model that approximates the behavior of the original system, there has been a significant interest in achieving additional desirable properties. In what follows, we summarize the most desirable properties along with related works in the literature regarding these properties.

\subsection{Literature Review}\label{subsec:properties}
In the following, we highlight some properties that are desired from a reduced order model. A notational remark: we denote an FOM of order $n$ by $\Sigma^n$ and its respective ROM of order $r$ by $\Sigma^r$.
\begin{enumerate}
    \item \textbf{Minimal Error between $\Sigma^r$ and $\Sigma^n$:} 
    A primary requirement from any model order reduction approach is that it yields a ROM which approximates the behavior of the FOM well in the sense of minimizing the approximation error measured with some metric. Generally, the $\mathcal{H}_2$, $\mathcal{H}_{\infty}$, or Hankel norm of the error are minimized, where the error is defined either as the difference between the outputs or the difference between the transfer functions of the two systems \cite{gugercin2008h_2,bryson1990second,774107,spanos1992new,shakib2021optimal,necoara2019parameter}.
    
    While many approaches that minimize the error have been proposed, most of these approaches are numerical, i.e., they do not have explicit theoretical guarantees (such as a provable upper bound on the error). Singular Value Decomposition (SVD) based approaches, such as  balanced truncation \cite{moore1981principal}, are well known to provide an explicit theoretical bound on the error. 
    Hankel norm approximation and singular perturbation approximation are two other SVD based approaches that have the same error bound as balanced truncation. We refer to the surveys \cite{antoulas2001survey,benner2015survey} for more details. 

    \item \textbf{Robustness to Modeling Errors:} Generally, the parameters of the state-space model are not exactly known due to the presence of modeling errors (which can be viewed as a special class of disturbances). Model reduction of uncertain systems has largely been focused on systems with specific structures, such as systems modeled by Linear Fractional Transform (LFT) representations \cite{beck1996model,li2010gramian,wang1991model,li2014coprime,beck2006coprime} and polytopic uncertain linear systems \cite{wu1996induced}. 


    \item \textbf{Preserving Stability:} A model reduction approach is said to preserve the stability of a stable system $\Sigma^n$ if it yields a system $\Sigma^r$ which is stable. In general, most existing approaches, with SVD based approaches being a notable exception, do not preserve stability. Some common techniques 
    include deleting unstable poles in post-processing \cite{bai1997stable,jaimoukha1997implicitly},  incorporating SVD based approaches \cite{yousefi2006preserving,selga2012stability}, or combining techniques from dissipativity theory \cite{ionescu2010moment,pulch2019stability}. These approaches are either numerical or do not satisfy other properties such as robustness.

    It is worth mentioning that the importance of an explicit theoretical bound on the error as well as the preservation of stability of balanced truncation has been widely recognized, owing to which balanced truncation is one of the most popular approaches for model order reduction. However, a major drawback of this approach is that it requires solving two Lyapunov equations which can be computationally expensive. Thus, these approaches may be suitable only for moderate scale settings. A popular approach that is suitable for large scale settings is moment matching; however, this approach does not preserve stability or provide an explicit error bound. Thus, there has been a recent interest in optimization centered approaches based on moment matching that minimize the error while ensuring stable ROMs \cite{necoara2022optimal,shakib2021optimal}.

    \item \textbf{Compositionality:} Many complex large-scale systems consist of interconnected high-dimensional subsystems. One approach is to reduce the entire interconnected system as a whole using any model order reduction technique. While feasible, this approach does not preserve the interconnection structure \cite{lutowska2012model}. Although structure preserving methods exist \cite{lutowska2012model,sandberg2009model,schilders2014novel,vandendorpe2008model,necoara2019h2}, they either lack error bounds or work only under restrictive settings. Modular model order reduction is an alternate and a popular approach in which each subsystem is approximated by its ROM. This approach preserves the interconnection structure and is computationally cheaper. However, in this approach, characterizing how approximation errors on a subsystem level affect the stability and accuracy of the interconnected system is challenging. Works such as \cite{ishizaki2013model} and \cite{cheng2019balanced} have characterized an error bound for bidirectional networks. More recently, using a robust performance analysis approach, \cite{janssen2024modular} characterizes an error bound on the interconnected system, given the approximation errors of the subsystems. We refer to \cite{cheng2021model} and the references therein for an in-depth review on this line of literature.

    \item \textbf{Ensuring Closed-Loop Stability of $\Sigma^n$:} Most works discussed so far consider approximation errors for the open-loop behavior. However, within the context of feedback systems, it is even more important to consider approximation errors in terms of the difference of the closed-loop behavior of the systems with the same controller. Gap metrics \cite{georgiousmith1990, vinnicombe1993frequency,vinnicombe2000uncertainty}, such as the gap and the $\nu$-gap, provide a measure of distance between open-loop systems in terms of their closed-loop behaviors. The fundamental property of gap metrics can be informally stated as follows. If a controller performs sufficiently well with a system $\Sigma_1^n$, and if the distance (in the sense defined by the gap metric) between system $\Sigma_1^{n_1}$ and $\Sigma_2^{n_2}$ is sufficiently small, then the same controller is guaranteed to achieve a certain level of performance with $\Sigma_2^{n_2}$. 
    Although the existence of a reduced model such that the gap or $\nu$-gap between the FOM and the ROM is less than a specified value  has long been established \cite{buskes2007reduced, cantoni2001model}, to the best of our knowledge, only numerical results exist to obtain a reduced model that satisfies such a gap metric error bound \cite{buskes2008step,sootla2013nu,eldesoukey2022observations}. 
    Since one of the applications of model order reduction is to design a controller for the original model, an explicit theoretical bound (as established in this work) on the output of the FOM with a controller designed using the ROM is clearly of interest.

\end{enumerate}

While much work exists concerning each of these desirable properties, there is still a growing interest in this area, as most of the works are experimental (i.e., lack theoretical guarantees) or require restrictive assumptions.


\subsection{Motivation}
Simulating the FOMs is crucial for analyzing the FOM or even designing control inputs in real-time \cite{schilders2008model}. For systems such as transportation networks, such simulations may be run online, i.e., in conjunction with the FOMs. Thus, one of the key applications of reduced-order models is to reduce the computational cost of simulations.
Given this application, our motivation is as follows: since an FOM may operate in the presence of arbitrary disturbances, its ROM must serve as a good approximation, even in the presence of unmodeled disturbances. We refer to this property, which is the primary motivation of this work, as \emph{robustness} of a ROM. 


\begin{figure}[t]
\begin{subfigure}{0.48\columnwidth}
  \centering
  \includegraphics[width=\linewidth]{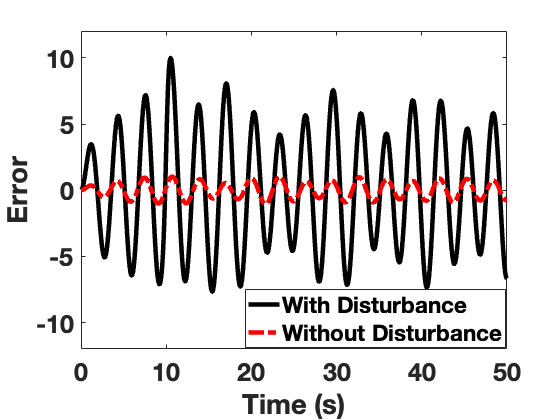}
 \caption{ROM obtained via balanced truncation.}
  \label{fig:BT_not_robust}
\end{subfigure}
\hfill
\begin{subfigure}{0.48\columnwidth}
  \centering
  \includegraphics[width=\linewidth]{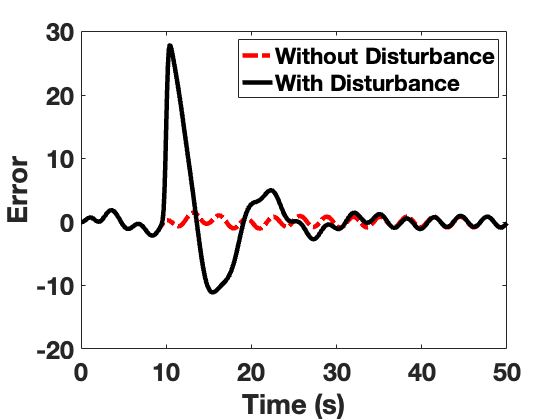}
 \caption{ROM obtained via moment-matching.}
  \label{fig:MM_not_robust}
\end{subfigure}
\caption{ Difference between the outputs (error) of the FOM and its ROM.
}
\label{fig:not_robust}
\end{figure}

Recall that balanced truncation and moment matching are two of the most popular methods for model order reduction. While we will discuss these methods in detail in Section \ref{sec:prelim}, to further motivate the need for frameworks that yield robust ROMs, we illustrate via Figure \ref{fig:not_robust} that balanced truncation and moment matching are not robust. 
In Figure \ref{fig:not_robust}, we consider two scenarios.
In the left plot, we consider an ROM obtained via balanced truncation. Specifically, we 
concatenate the system matrices associated with the input and the disturbance and then apply balanced truncation on the modified system. It must be highlighted that this approach is not suitable for the aforementioned application as the disturbance is treated as an input and thus requires the disturbance to be known. Alternatively, one may choose to ignore the disturbance. 
For this case, we construct an ROM via moment matching of the given FOM ignoring the disturbance.
For both cases, as Figure \ref{fig:not_robust} illustrates, while the error is small in the absence of a disturbance, it is much larger in the presence of a disturbance. In other words,   the ROM obtained through moment matching or balanced truncation is not robust. We remark that, for balanced truncation, the error in Figure \ref{fig:not_robust} is within the known error bound (refer to Section \ref{sec:prelim} for error bound).
Further, balanced truncation is the only method with an analytical error bound.
This means that the (only known) error bound may be conservative, especially in the presence of a disturbance.

Most of the prior work on robust model order reduction has either only focused on robustness to modeling errors or required restrictive assumptions on the disturbance. While these results provide valuable insights, they are not applicable for arbitrary disturbances.
Recently in \cite{pirastehzad2024comparison}, a perturbation based approach was proposed to force a linear system to \emph{behave similarly} to another linear system, for arbitrary inputs and disturbances. Motivated by \cite{pirastehzad2024comparison},  we adopt a \emph{similarity} based approach for robust model order reduction. We summarize our specific contributions in the next subsection.

\subsection{Contributions}\label{subsec:contributions}
In this work, for continuous-time linear systems, we provide a framework for a robust model order reduction, i.e., the output of the obtained ROM continues to remain a good approximation of the FOM, even in the presence of arbitrary disturbances. Additionally, our framework satisfies all of the properties described in Section \ref{subsec:properties}. Specifically, in this work, we provide a framework for model order reduction which: 
\begin{enumerate}
    \item minimizes the error (defined as the $\mathcal{L}_2$-norm between the output of the full and the reduced systems) and provides an analytical upper bound on the error;

    \item is robust to arbitrary $\mathcal{L}_2$ disturbances;

    \item preserves the stability of the original system;
    
    \item provides an explicit error bound between an arbitrary interconnected system and its reduced model while preserving the interconnection structure; and
    
    \item provides an upper bound on the (closed-loop) 
    output of the original system when a controller designed using the reduced model is used with the original system;
    
    \item is compatible with existing approaches such as balanced truncation and moment matching, thereby extending some of its properties to those approaches as well (while inheriting desirable features of those approaches).
\end{enumerate}

To the best of our knowledge, this is the first framework to satisfy all of the aforementioned six properties.
The rest of this work is organized as follows. In Section \ref{sec:prelim}, we review some required concepts and approaches and formally state the problem. In Section \ref{sec:MOR}, we describe our approach and establish that it satisfies the properties (1)-(3) described in Section \ref{subsec:contributions}. In Section \ref{sec:compositionality} and Section \ref{sec:closed_loop} we establish the compositionality and closed-loop properties, respectively. A description of how the proposed framework can be used in conjunction with existing approaches is provided in Section \ref{sec:Add_on}. In Section \ref{sec:numerics}, we numerically establish the efficiency of our approach. Finally, a summary of this work and an outline of future directions is given in Section \ref{sec:conclusion}.

\textit{Notation:} For a matrix $M\in \mathbb{R}^{n\times n}$, we denote its spectrum and the largest (resp. smallest)  eigenvalue as $\text{spec}(M)$ and $\lambda_{\text{max}}(M)$ (resp. $\lambda_{\text{min}}(M)$), respectively. We use $\textbf{I}$ and $\textbf{0}$ to denote an identity and a zero matrix, respectively, of appropriate dimensions. For a symmetric matrix, the symmetric terms are denoted by $*$.
For any two vectors $x_1\in\mathbb{R}^{n_1}$ and $x_2\in\mathbb{R}^{n_2}$, we define the operator $\text{col}(\cdot , \cdot)$ such that $\text{col}(x_1,x_2) = \left(x_1^{\top},x_2^{\top} \right)^{\top}$. Given a vector $x\in\mathbb{R}^n$, $\abs{x}\coloneqq \left(x^{\top}x \right)^{1/2}$.
We denote by $\mathcal{L}_2$ the space of measurable functions $u:[0,\infty)\to\mathbb{R}^n$ such that $\int_0^\infty \abs{u(t)}^2\mathrm{d}t<\infty$. Accordingly, the $\mathcal{L}_2$ space is endowed with the norm $\norm{u}^2 = \int_0^{\infty}\abs{u(t)}^2\mathrm{d}t$. Given a matrix $M$, $M\succ 0$ (resp. $M\prec 0$) denotes that $M$ is positive (resp. negative) definite. Similarly, $M\succeq 0$ (resp. $M\preceq 0$) denotes that $M$ is positive (resp. negative) semi-definite. 
Given a matrix $M\succ 0$, $\abs{x}_M\coloneqq \left(x^{\top}Mx \right)^{1/2}$ and
 $\norm{u}_M^2\coloneqq \int_0^\infty \abs{u(t)}_M^2\mathrm{d}t$.

\section{Preliminaries, Motivation, and Problem Statement}
We begin by reviewing two of the most popular approaches for model order reduction, namely, balanced truncation and moment matching, as well as basic concepts regarding the similarity of two linear systems. We then formally describe the problem considered in this work.


\subsection{Preliminaries}\label{sec:prelim}
To review the balanced truncation and the moment matching methods, we consider a linear continuous-time system
\begin{align}\label{eq:dynamics_BT}
    \Sigma^{n} \colon \begin{cases}
        \dot{x} &= Ax + Bu;\\
        y &= Cx, 
    \end{cases}
\end{align}
where $x \in \mathbb{R}^{n}$, $u\in \mathbb{R}^{m}$, and $y\in\mathbb{R}^{p}$ represent the system state, input, and output, respectively. Throughout this section, we assume that $A$ is Hurwitz.

\subsubsection{Balanced Truncation}

Balanced truncation (BT) is the most well-known model reduction method. For a given linear continuous-time system of the form \eqref{eq:dynamics_BT}, balanced truncation requires solving the following two Lyapunov equations,
\begin{equation*}
    A\mathcal{C} + \mathcal{C}A^{\top} + BB^{\top}=0;~ A\mathcal{O} + \mathcal{O}A^{\top} + CC^{\top}=0,
\end{equation*}
where matrices $\mathcal{C}$ and $\mathcal{O}$ are the reachability and observability Gramians, respectively.
Under the assumption that the matrix $A$ is Hurwitz, both $\mathcal{C}$ and $\mathcal{O}$ are positive semideﬁnite matrices. The square root of the eigenvalues of $\mathcal{C}\mathcal{O}$ are the singular values of the Hankel operator associated with the system \eqref{eq:dynamics_BT} and are known as \emph{Hankel singular  values} denoted as $\sigma_i^H$, $1\leq i\leq n$. For ease of exposition, assume that all the Hankel singular values are distinct. Then, a reduced model via balanced truncation is obtained by removing the 
states that correspond to `sufficiently small' Hankel singular values. Specifically, given a positive integer $r<n$, a reduced model $A_r$, $B_r$, and $C_r$ of order $r$ is obtained via balanced truncation as follows. First, the Cholesky factors $L_{\mathcal{C}}$ and $L_{\mathcal{O}}$ of $\mathcal{C}$ and $\mathcal{O}$, respectively, are computed. Then, the singular value decomposition of $L_{\mathcal{O}}^{\top}L_{\mathcal{C}}$ is determined by equating $L_{\mathcal{O}}^{\top}L_{\mathcal{C}} = ZSY$, where $S\coloneqq \text{diag}(\sigma_1^H,\dots,\sigma_n^H)$ and $Z$ (resp. $Y$) denote the left (resp. right) singular vectors. 
Let $S_r\coloneqq \text{diag}(\sigma_1^H,\dots,\sigma_r^H)$ and define 
$W_{\text{BT}} = L_{\mathcal{O}}Y_rS_r^{-1/2}$ and $V_{\text{BT}} = L_{\mathcal{C}}Z_rS_r^{-1/2}$, where $Z_r$ (resp. $Y_r$) denotes the leading $r$ columns of $Z$ (resp. $Y$). The reduced model is then given by projection, i.e., $A_r = W_r^{\top}AV_r$, $B_r = W_r^{\top}B$, and $C_r = CV_r$.
Balanced truncation has the following properties:
\begin{enumerate}
    \item The reduced system is asymptotically stable.
    \item The $\mathcal{H}_{\infty}$-norm, defined as $\norm{\Sigma}_{\mathcal{H}_{\infty}}\coloneqq \sup_{w\in\mathbb{R}}|\Sigma(\iota w)|$, of the error system satisfies
    \begin{align*}
        \norm{\Sigma-\Sigma_r}_{\mathcal{H}_\infty} \leq 2(\sigma_{r+1}^H+\dots +\sigma_{n}^H).
    \end{align*}
\end{enumerate}
Recently,~\cite{redmann2020lt2} established a similar bound for balanced truncation in the time-domain. Formally, the $\mathcal{L}_2$ bound on the error between the outputs satisfies
\begin{align}\label{eq:bound_BT}
        \norm{y-y_r}^2 \leq 2(\sigma_{r+1}^H+\dots +\sigma_{n}^H) \norm{u}^2,
    \end{align}
where $y_r$ denotes the output of the ROM.
\subsubsection{Moment Matching}
For large-scale systems, another popular method for model order reduction is moment matching in which the transfer function (or some derivative of it) of the obtained reduced model approximately matches that of the original model at specified frequencies. A $0$-moment of system \eqref{eq:dynamics_BT} at $s^* \in \mathbb{C}\setminus \text{spec}(A)$ is the transfer function of $\Sigma$ at $s^*$ \cite{astolfi2010model}. Let $s_1,\dots, s_r\in \mathbb{R}$ be given and let matrices $S\in\mathbb{R}^{r\times r}$ and $L\in\mathbb{R}^{r\times m}$ be fixed such that the pair $(L,S)$ is observable. In particular, the matrix $S$ is chosen such that $\text{spec}(S)=\{s_1,\dots,s_r\}$. Further, let $T\in \mathbb{R}^{n\times r}$ be the solution to the Sylvester equation:
\begin{align*}
    AT + BL = TS.
\end{align*}
Then, the reduced system $\Sigma^r$, parameterized by $B_r$ can be obtained as
\begin{align}\label{eq:family_MM}
    A_r\coloneqq S-B_rL, \quad C_r\coloneqq CT.
\end{align}
Equation \eqref{eq:family_MM} describes a family of reduced order 
models, each of order $r$, that achieve moment matching at frequencies specified by $\text{spec}(S)$. Note that the family of reduced models are parameterized by $B_r$. Further, the reduced system $\Sigma^r$ satisfies $\text{spec}(S-B_rL)\cap \text{spec}(S)=\emptyset$. It is known that moment matching, in general, does not guarantee that the ROM is stable even if the FOM is stable or provide an explicit error bound on how much the systems differ. Recent approaches focus on minimizing the $\mathcal{H}_2$ or $\mathcal{H}_{\infty}$-norm of the error by posing the problem as an optimization problem. 
In these approaches, the parametric freedom in the method is exploited, i.e., the parametrization is combined with an objective function (such as $\mathcal{H}_2$-norm). The reduced order model then obtained by minimizing the objective function with respect to $B_r$ minimizes the error between the systems. Since matrices $S$ and $L$ are fixed a priori, one may select them to ensure that the stability of the original model is preserved. We refer to \cite{necoara2022optimal,shakib2021optimal,astolfi2010model} for more details.

Next, we review a framework introduced in~\cite{pirastehzad2024comparison} that characterizes the \emph{similarity} of two linear systems.

\subsubsection{$(\gamma,\delta)$-Similar Systems}
Consider continuous-time linear systems of the form
\begin{equation}\label{eq:dynamics_prelim}
    \Sigma_i^{n_i} \colon \begin{cases}
        \dot{x_i} &= A_ix_i + B_iu_i + E_id_i;\\
        y_i &= C_ix_i, 
    \end{cases}
\end{equation}
where $x_i \in \mathbb{R}^{n_i}$, $u_i\in \mathbb{R}^{m}$, $d_i\in \mathbb{R}^{q_i}$, and $y_i\in\mathbb{R}^{p}$ represent the system state, input, disturbance, and output, respectively. Assume that the systems $\Sigma_i^{n_i}$ are $0$-asymptotically stable, i.e., the system is asymptotically stable in the absence of input and disturbance (equivalently, the system matrices $A_i$ are Hurwitz). 

\begin{definition}\label{def:similar_systems}
   Given two systems $\Sigma_i^{n_i}$, $i\in\{1,2\}$, of the form \eqref{eq:dynamics_prelim} and positive constants $\gamma$ and $\delta$, the system $\Sigma_2^{n_2}$ is said to be $(\gamma,\delta)$-similar to $\Sigma_1^{n_1}$, if there exist constants $\epsilon,\eta,\mu>0$ such that for every input $u_1,u_2\in \mathcal{L}_2$ and disturbance $d_1\in\mathcal{L}_2$, there exists a disturbance $d_2\in\mathcal{L}_2$ such that
   \begin{equation}\label{eq:similarity}
   \begin{split}
       \norm{y_1-y_2}^2\leq & \gamma\norm{u_1-u_2}^2 + (\delta-\epsilon)\norm{\text{col}(u_1,u_2)}\\
       &+ (\mu-\epsilon)\norm{d_1}^2 - \eta\norm{d_2}^2.
   \end{split}
   \end{equation}
\end{definition}
\medskip
The notion of $(\gamma,\delta)$-similarity in Definition \ref{def:similar_systems} measures the similarity of the trajectories of $\Sigma_1^{n_1}$ and $\Sigma_2^{n_2}$ in terms of their input-output behavior. The parameter $\gamma$ characterizes the deviation in the outputs with respect to the dissimilarity in the inputs.  The parameter $\delta$ characterizes the deviation in the outputs with respect to the individual inputs. Finally, the constants $\mu$ and $\eta$ characterize the deviation in the output with respect to disturbances. 
The notion of $(\gamma,\delta)$-similarity has the following property \cite[Proposition 1]{pirastehzad2024comparison}.
\begin{lemma}\label{lem:prelim_system_iteslef}
    There exists a $\gamma>0$ such that the system $\Sigma^n$ is $(\gamma,\delta)$-similar to itself for any $\delta>0$.
\end{lemma}

Determining the disturbance $d_2\in\mathcal{L}_2$ and the constants $\epsilon,\eta,\mu>0$ from Definition \ref{def:similar_systems} can be challenging. However, an algebraic characterization of Definition \ref{def:similar_systems} was established in \cite{pirastehzad2024comparison} as follows. Let $x\coloneqq \text{col}(x_1,x_2)$, $w\coloneqq \text{col}(u_1,u_2,d_1)$, and $z\coloneqq \text{col}(y_1-y_2,d_2)$ and consider the following composite system obtained by collecting the dynamics of $\Sigma_1^{n_1}$ and $\Sigma_2^{n_2}$:
\begin{equation}\label{eq:composite_system}
\begin{split}
    \dot{x} &= Ax + Bd_2 + Ew,\\
    &z = Cx + Dd_2,
\end{split},
\end{equation}
where
\begin{equation}\label{eq:composite_matrices_prelim}
\begin{split}
    &A = \begin{bmatrix}
        A_1 & \mathbf{0}\\ \mathbf{0} & A_2
    \end{bmatrix},
    B = \begin{bmatrix}
        \mathbf{0} \\ E_2
    \end{bmatrix}, E = \begin{bmatrix}
        B_1 & \mathbf{0} & E_1\\ \mathbf{0} & B_2 & \mathbf{0}
    \end{bmatrix},\\
    &C = \begin{bmatrix}
        C_1 & -C_2\\ \mathbf{0} & \mathbf{0}
    \end{bmatrix}, D = \begin{bmatrix}
        \mathbf{0} \\ \mathbf{I}
    \end{bmatrix}.
\end{split}
\end{equation}
Further, let
\begin{equation}\label{eq:Q_and_R_prelim}
    Q\coloneqq \begin{bmatrix}
    (\gamma+\delta)\mathbf{I} & -\gamma\mathbf{I} & \mathbf{0}\\
    -\gamma\mathbf{I} & (\gamma+\delta)\mathbf{I} & \mathbf{0}\\
    \mathbf{0} & \mathbf{0} & \mu\mathbf{I}
    \end{bmatrix}, \quad
    R = \begin{bmatrix}
        \mathbf{I} & \mathbf{0}\\
        \mathbf{0} & \eta\mathbf{I}
    \end{bmatrix}.
\end{equation}

Then the following result, established in \cite[Theorem 2]{pirastehzad2024comparison}, characterizes an algebraic necessary and sufficient condition for the notion of $(\gamma,\delta)$-similarity of two systems.
\begin{theorem}\label{thm:LMI_prelim}
    For $\gamma,\delta>0$, $\Sigma_2^{n_2}$ is $(\gamma,\delta)$-similar to $\Sigma_1^{n_1}$ if and only if there exist a positive definite matrix $X$, a matrix $\Pi$, and positive scalars $\eta$ and $\mu$ such that
    \begin{equation}\label{eq:LMI_prelim}
    \begin{split}
        &\begin{bmatrix}
        AX + B\Pi + (AX+B\Pi)^\top & * & *\\
        E^\top & -Q(\gamma,\delta) & * \\
        CX+D\Pi & \mathbf{0} & -R
    \end{bmatrix}\prec 0.
    \end{split}
    \end{equation}
\end{theorem}
It is established in \cite[Lemma 3]{pirastehzad2024comparison} that the disturbance $d_2$ that must be applied to $\Sigma_2^{n_2}$ can be chosen in a closed-loop fashion, i.e., as a static state feedback which depends of the state of both the systems. Mathematically, $d_2 = Fx$, where $F\coloneqq \Pi X^{-1}$. Further, the following result, established in \cite[Proposition 4]{pirastehzad2024comparison}, will be instrumental in establishing some of the results in this work.

\begin{lemma}\label{lem:prelim_output_bound}
    Suppose, for some $\gamma,\delta>0$, $\Sigma_2^{n_2}$ is $(\gamma,\delta)$-similar to $\Sigma_1^{n_1}$, where $\Sigma_1^{n_1}$ and $\Sigma_2^{n_2}$ are $0$-asymptotically stable. Then, there exist positive constants $l$ and $k$ such that, for all $u_1,u_2, d_1\in \mathcal{L}_2$,
    \begin{align*}
    \norm{\text{col}(y_1,y_2}^2\leq l\norm{\text{col}(u_1,u_2}^2
        + k\norm{d_1}^2.
    \end{align*}
\end{lemma}



\subsection{Problem Statement}\label{subsec:problem}
We are interested in solving the following problem.
\begin{problem}\label{prob:1}
Given a system $\Sigma^n$ as described in \eqref{eq:dynamics_prelim} and a positive integer $r<n$, determine a reduced order model $\Sigma^r$ which satisfies the following:
    \begin{enumerate}
    \item $||y - y_r||$ is minimized;
    \item $\Sigma^r$ is stable;
    \item $\Sigma^r$ is robust to any unknown disturbances or modeling errors $d_1\in\mathcal{L}_2$ that may be applied to $\Sigma^n$.
\end{enumerate}
Further, if such a system $\Sigma^r$ exists, then:
\begin{enumerate}
\setcounter{enumi}{3}
    \item for a given arbitrary interconnected system $\Sigma^n$ consisting of $N$ subsystems $\Sigma^{n_i}_i, ~1\leq i\leq N$ , determine an ROM $\Sigma^r$ of $\Sigma^n$, such that the interconnection structure is preserved. Further, characterize an upper bound on the error between the system $\Sigma^n$ and its reduced model $\Sigma^r$.
    \item Given a stabilizing controller for $\Sigma^r$, characterize an upper bound on the output of the closed-loop system when the same controller is used for $\Sigma^n$.
\end{enumerate}
\end{problem}

In what follows, we provide a general framework based on the concepts of $(\gamma,\delta)$-similarity of systems (see Section \ref{sec:prelim}) that can be used to solve Problem \ref{prob:1}. Further, as we show in Section \ref{sec:Add_on}, the proposed framework can also be used as an \emph{add-on} with existing approaches such as balanced truncation and moment matching, thereby extending some of its properties to those approaches as well, while inheriting desired features of those approaches.

\section{$(\gamma,\delta)$-Model Order Reduction}\label{sec:MOR}
Observe that the notion of $(\gamma,\delta)$-similarity between two systems is independent of the respective orders of the two systems  as it characterizes the similarity between two systems only in terms of their input-output behavior. Given this observation, our approach can be summarized as follows; given a system $\Sigma^n$ and a positive integer $r<n$, we aim to determine a reduced order model $\Sigma^r$ and a suitable perturbation $d_2\in\mathcal{L}_2$ such that the obtained ROM is $(\gamma,\delta)$-similar to $\Sigma^n$. Note that the disturbance $d_{2}$ will depend on the state of both the ROM and the FOM. This is unavoidable if there is no a priori structure or assumption that we can impose on the disturbance $d_{1}$ affecting the original system. Such requirements have been imposed in the past for model order reductions \cite{samari2025model, kurtz2019formal}.
In the sequel, the system $\Sigma^r$ which is thus determined and is $(\gamma,\delta)$-similar to $\Sigma^n$ by construction will be referred to as a $(\gamma,\delta)$-ROM of $\Sigma^n$. 
We characterize this formally below.
\begin{definition}\label{def:gamma_delta_ROM}
    Given $0$-asymptotically stable systems $\Sigma^n$ and $\Sigma^r$ with $r<n$, $\Sigma^r$ is said to be a $(\gamma,\delta)$-ROM of $\Sigma^n$, if there exist positive constants $\gamma,\delta,\epsilon,\eta,\mu$ and a \emph{driving input} $d_2\in\mathcal{L}_2$ such that $\forall u_1,u_2,d_1\in\mathcal{L}_2$
   \begin{equation}\label{eq:similarity_ROM}
   \begin{split}
       \norm{y_1-y_2}^2\leq & \gamma\norm{u_1-u_2}^2 + (\delta-\epsilon)\norm{
           \text{col}(u_1,u_2)
       }^2 \\
       &+ (\mu-\epsilon)\norm{d_1}^2 - \eta\norm{d_2}^2. 
   \end{split}
   \end{equation}
\end{definition}
Observe that the notion of $(\gamma,\delta)$-ROM requires determining the positive constants $(\gamma,\delta)$ as opposed to the constants given a priori when we need to analyze whether two systems are $(\gamma,\delta)$-similar. 
Further, Definition \ref{def:gamma_delta_ROM} requires that a reduced model $\Sigma^r$ be given a priori. This suggests that the notion of $(\gamma,\delta)$-ROM may work in conjunction with the existing approaches. 
Although true (we will consider this scenario more formally in Section \ref{sec:Add_on}), we highlight that this is not the only way to utilize the notion of $(\gamma,\delta)$-ROM. 

In what follows, we will first provide a framework which, for a given order $r<n$, yields a $(\gamma,\delta)$-ROM $\Sigma^r$ of $\Sigma^n$. 
We will also establish that the obtained $\Sigma^r$ satisfies all the desirable properties described in Problem~\ref{prob:1}.

From Definition \ref{def:gamma_delta_ROM} and Theorem \ref{thm:LMI_prelim}, we begin by recasting Problem \ref{prob:1} as an optimization problem, which we will refer to as $\mathcal{P}_0$.
Consider a FOM $\Sigma^n$ of the form \eqref{eq:dynamics_prelim}. Then, the optimization problem $\mathcal{P}_0$ is the following:
\begin{align}
&\textrm{Problem $\mathcal{P}_{0}$: }    \min_{A_2,B_2,C_2,X\succ 0,\Pi,\gamma>0,\delta>0} \left(\gamma+ \delta\right) \nonumber\\
    &\text{subject to} \nonumber\\
    &\qquad  \sigma(\Sigma_r)\subset \mathbb{C}^- \label{eq:stability_constraint}\\
    & \begin{bmatrix} \label{eq:BMI_constraint}
        AX + B\Pi + (AX+B\Pi)^\top & * & *\\
        E^\top & -Q(\gamma,\delta) & * \\
        CX+D\Pi & \mathbf{0} & -R
    \end{bmatrix} \prec 0,
\end{align}
where the matrices $A,B,C,$ and $D$ are defined in \eqref{eq:composite_matrices_prelim} and matrices  $Q$ and $R$ are defined in \eqref{eq:Q_and_R_prelim}.
Note that, since $A_2, B_2,$ and $C_2$ are unknown, the constraint specified in equation \eqref{eq:BMI_constraint} 
is a Bilinear Matrix Inequality (BMI) constraint. Further, the constraint defined in \eqref{eq:stability_constraint} ensures that the obtained ROM $\Sigma^r$ is stable.
Due to the presence of the stability constraint in \eqref{eq:stability_constraint}, the proposed optimization problem $\mathcal{P}_0$ is challenging to solve. The following result establishes that, for a given $0$-asymptotically stable system $\Sigma^n$, the obtained $(\gamma,\delta)$-ROM $\Sigma^r$ is stable if the BMI constraint in \eqref{eq:BMI_constraint} holds and thus we can drop the constraint  \eqref{eq:stability_constraint}. 

\begin{theorem}\label{thm:stability_preserved}
    Suppose that the systems $\Sigma^n$ and $\Sigma^r$ are such that equation \eqref{eq:BMI_constraint} holds. Further, let $\Sigma^n$ be $0$-asymptotically stable. Then, $\Sigma^r$ is $0$-asymptotically stable.
\end{theorem}
\begin{proof}
Let $F:=\Pi X^{-1}$, $A_F := A+BF$, and $C_F:=C+DF$. Then, taking congruence transformation of  \eqref{eq:BMI_constraint} with respect to 
    \begin{align*}
        \begin{bmatrix}
            X^{-1} & \mathbf{0} & \mathbf{0}\\
            \mathbf{0} & \mathbf{I} & \mathbf{0}\\
            \mathbf{0} & \mathbf{0} & \mathbf{I}
        \end{bmatrix}
    \end{align*}
    yields the constraint
    \begin{align}\label{eq:LMI_transf_2.1}
        \begin{bmatrix}
            A_F{^\top} X^{-1} + X^{-1}A_F & X^{-1}E & C_F{^\top}\\
            E^\top X^{-1} & -Q & \mathbf{0}\\
            C_F & \mathbf{0} & -R
        \end{bmatrix}\prec 0.
    \end{align}
    Since $X\succ 0$, it follows that $A_F$ is Hurwitz. For $\tilde{R}\coloneqq R^{-1}$ and taking the Schur complement of \eqref{eq:LMI_transf_2.1} yields
    \begin{align*}
        \begin{bmatrix}
            A_F^{\top}X^{-1} + X^{-1}A_F + C_F^{\top}\tilde{R}C_F & X^{-1}E\\
            E^{\top}X^{-1} & -Q
        \end{bmatrix} \prec 0.
    \end{align*}
    This implies that the composite system defined in equation \eqref{eq:composite_system} with $d_2=Fx$ is dissipative with the respect to the supply rate 
    \begin{align}\label{eq:supply_rate_stability}
        s(w,z) = 
        \begin{bmatrix}
            w\\z
        \end{bmatrix}^{\top}\begin{bmatrix}
            Q & \mathbf{0}\\
            \mathbf{0} & -\tilde{R}
        \end{bmatrix}\begin{bmatrix}
            w\\z
        \end{bmatrix}
    \end{align}
    and storage function $V(x)=x^\top X^{-1}x$ \cite[Theorem 2.9]{scherer2000linear}. From the definition of a dissipative system,
    \begin{align*}
        V(x(t_1)) \leq V(x(t_0)) + \int_{t_0}^{t_1}s(w(t),z(t))dt-\epsilon\int_{t_0}^{t_1}|w(t)|^2dt,
    \end{align*}
    for all $t_0\leq t_1$. For $t_0=0$ and using equation \eqref{eq:supply_rate_stability}, we obtain
    \begin{align*}
        &\norm{z}^2_R-\norm{w}_Q^2 \leq V(x(0)) - V(x(t_1)) -\epsilon\norm{w}^2\\
        &\implies \norm{z}^2_R \leq \norm{w}_Q^2 + V(x(0)) -\epsilon\norm{w}^2\\
        &\implies \norm{z}_R \leq \sqrt{\lambda_{\text{max}}\left(Q-\epsilon \right)}\norm{w}+ \sqrt{V(x(0))}.
    \end{align*}
    Recall that $z(t)\coloneqq z(t,0,x(0),d,w)$.
    Since the composite system is linear, it follows that $z(t) = z(t,0,x(0),0,0)+z(t,0,0,d,w)$. Then, using triangle inequality yields
    \begin{multline}\label{eq:output_bound_2.1}
        ||z(t,0,x(0),0,0)||_R \leq  \sqrt{\lambda_{\text{max}}\left(Q-\epsilon \right)}||w||+ \sqrt{V(x(0))} \\
         + ||z(t,0,0,d,w)||_R.
    \end{multline}
    Now, using equation \eqref{eq:supply_rate_stability} and from the definition of dissipativity, we obtain that for $t_0 =0$ and $x(0)=0$, $||z(t,0,0,d,w)||_R\leq \sqrt{\lambda_{\text{max}}\left(Q-\epsilon \right)}||w||$. Substituting in equation \eqref{eq:output_bound_2.1} yields 
    \begin{align*}
        \norm{z(t,0,x(0),0,0)}_R \leq & \sqrt{\lambda_{\text{max}}\left(Q-\epsilon \right)}\norm{w}\left(1+\frac{1}{\lambda_{\text{min}}\left( R \right)} \right)\\
        & + V(x(0)).
    \end{align*}
    This means that the $\mathcal{L}_2$ norm of the output is bounded under $d_2(t)=0$ and $w(t)=0$ for all $t$ implying that matrix $A$ is Hurwitz \cite[Theorem 3.1.1]{green2012linear}. 
    Since $A_1$ is Hurwitz (as $\Sigma^n$ is assumed to be $0$-asymptotically stable), it follows that $A_2$ must be Hurwitz. 
\end{proof}
As a consequence of Theorem \ref{thm:stability_preserved}, we can remove the stability constraint defined in equation \eqref{eq:stability_constraint} and reformulate the optimization problem $\mathcal{P}_0$ to the following optimization problem which we will refer to as $\mathcal{P}_1$.
\begin{align}
    &\textrm{Problem $\mathcal{P}_{1}$: } \min_{A_2,B_2,C_2,X\succ 0,\Pi,\gamma>0,\delta>0} \left(\gamma+\delta\right)\\
    &\text{subject to} \nonumber\\
    & \begin{bmatrix}
        AX + B\Pi + (AX+B\Pi)^\top & * & *\\
        E^\top & -Q(\gamma,\delta) & * \\
        CX+D\Pi & \mathbf{0} & -R
    \end{bmatrix} \prec 0.\nonumber
\end{align}

We now establish the main result of this section.

\begin{theorem}\label{thm:generalized_ROM_stable}
    Given a $0$-asymptotically stable system $\Sigma^n$ and a positive integer $r<n$, suppose that a solution $\{A_2^*,B_2^*,C_2^*,X^*,\Pi^*,\gamma^*,\delta^*\}$ to the optimization problem $\mathcal{P}_1$ exists. Then, the system $\Sigma^r$ defined by the system matrices $A_2^*,B_2^*,C_2^*$ is a $(\gamma^*,\delta^*)$-ROM of $\Sigma^n$. Further, $\Sigma^r$ is $0$-asymptotically stable.
\end{theorem}
\begin{proof}
    The result follows directly from the definition of $(\gamma^*,\delta^*)$-ROM and Theorem \ref{thm:stability_preserved}. 
\end{proof}

\begin{remark}
    Suppose that the solution to the optimization problem $\mathcal{P}_1$ exists. Then, the obtained ROM satisfies properties (1), (2), and (3) defined in Problem \ref{prob:1}.
\end{remark}

\begin{remark}
    Problem $\mathcal{P}_1$ consists of a BMI constraint. Many approaches for model order reduction require solving an optimization problem with BMI constraints (see, for instance, \cite{necoara2019parameter,necoara2022optimal,shakib2021optimal}). By imposing additional constraints on the matrix $X$ (similar to, for instance, in \cite{necoara2022optimal,ebihara2004h}), it might be possible to relax the BMI to a Linear Matrix Inequality (LMI) constraint. We leave this direction as future work. Further, in Section \ref{sec:Add_on}, we will see that when we combine our framework with balanced truncation, the BMI constraint automatically converts to an LMI constraint.
\end{remark}

Theorem \ref{thm:generalized_ROM_stable} ensures that three out of the five properties described in Section \ref{subsec:problem} are satisfied by the proposed approach. In what follows, we will establish that the proposed approach also satisfies properties (4)-(5) defined in Problem \ref{prob:1}.


\section{Model Order Reduction of Interconnected Systems}\label{sec:compositionality}

\begin{figure}[t]
\begin{subfigure}{0.45\columnwidth}
  \centering
  \includegraphics[width=\linewidth]{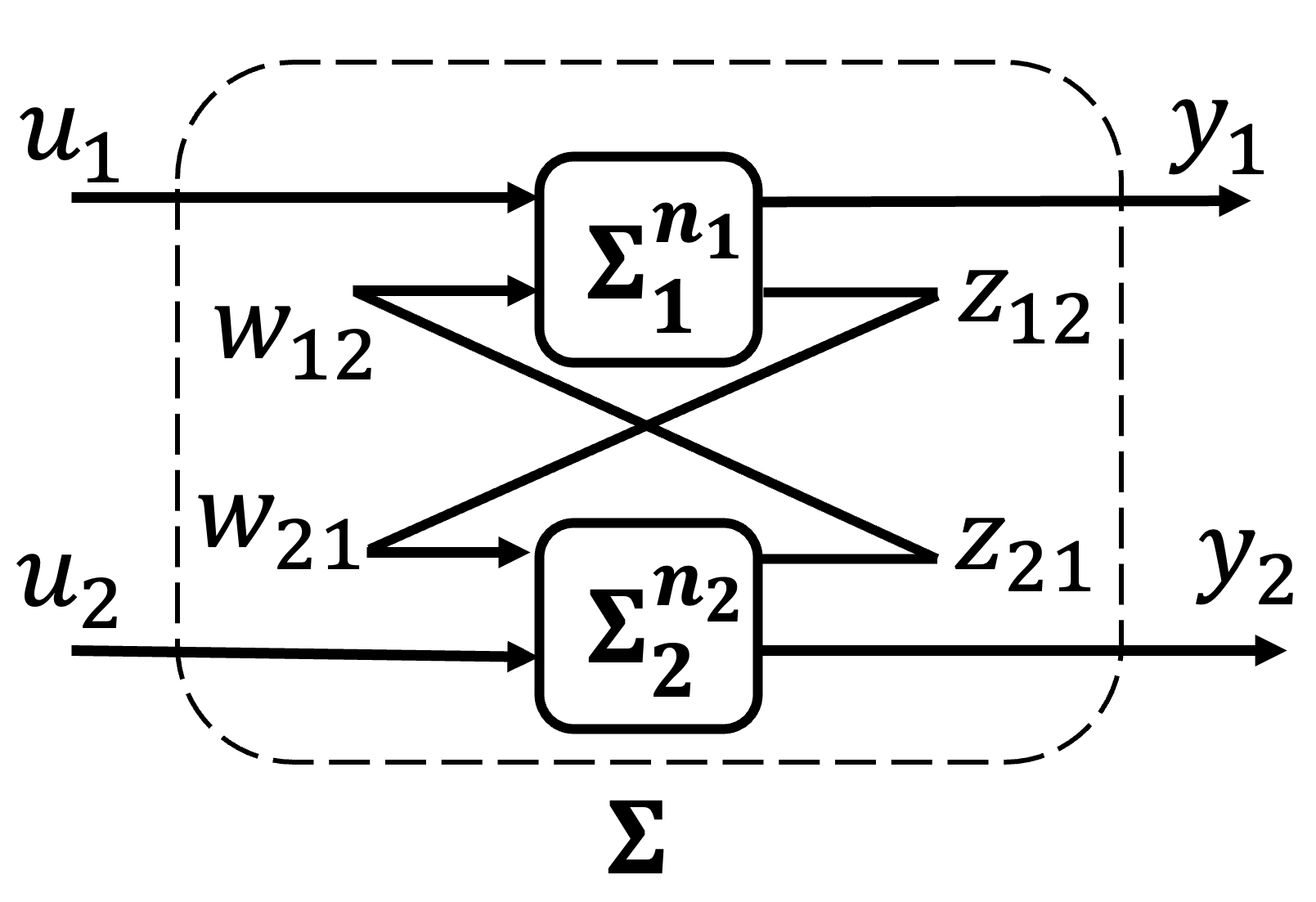}
  \caption{System $\Sigma$.}
  \label{fig:modular_FOM}
\end{subfigure}
\hfill
\begin{subfigure}{0.45\columnwidth}
  \centering
  \includegraphics[width=\linewidth]{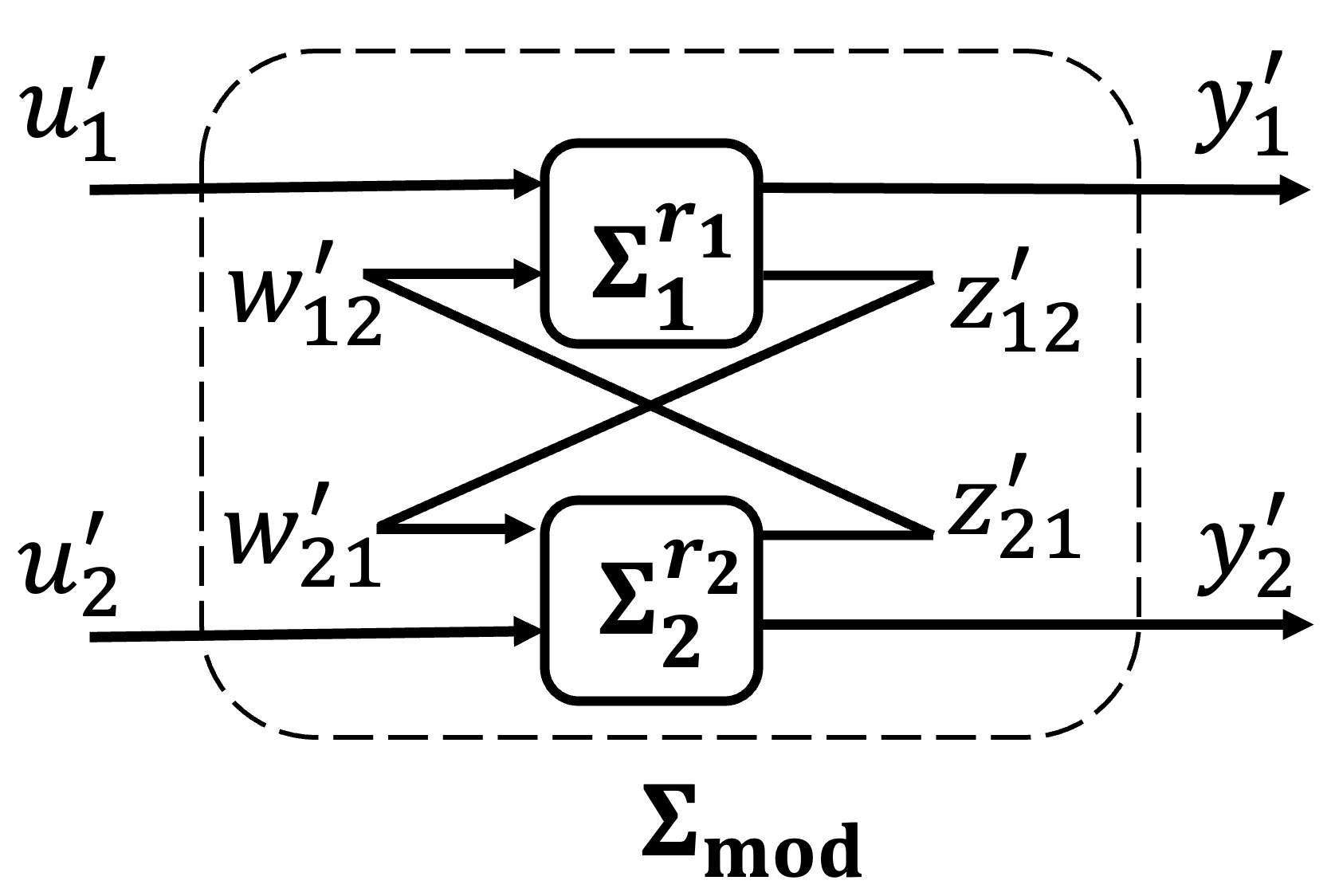}
  \caption{System $\Sigma_{\text{mod}}$.}
  \label{fig:modular_ROM}
\end{subfigure}
\caption{Illustration of the modular approach for system $\Sigma$ consisting of $N=2$ subsystems. Subsystem $\Sigma_1^{r_1}$ and $\Sigma_2^{r_2}$ are ROMs of $\Sigma_1^{n_1}$ and $\Sigma_2^{n_2}$, respectively. $w_{12}, w_{21}$ (resp. $w_{12}', w_{21}'$) and $z_{12}, z_{21}$ (resp. $z_{12}', z_{21}'$) are the internal inputs and outputs, respectively, to system $\Sigma$ (resp. $\Sigma_{\text{mod}}$). $u_1,u_2$ (resp. $u_1', u_2'$) and $y_1, y_2$ (resp. $y_1', y_2'$) are the external inputs and outputs, respectively, to system $\Sigma$ (resp. $\Sigma_{\text{mod}}$).}
\label{fig:modular}
\end{figure}

In this section, with the aim of addressing desirable property (4) in Problem \ref{prob:1}, we  consider interconnected systems consisting of $N>1$ subsystems, where each subsystem may have a high dimension. As a consequence, the interconnected system is also high-dimensional. As discussed earlier, a popular approach, referred to as a \emph{modular approach}, is to apply model-order reduction on the subsystem level and connect the resulting reduced order models of the subsystems (see Figure \ref{fig:modular}). A key challenge in the literature is to quantify how the accuracy of the resulting interconnected system of the subsystem ROMs is affected by the approximation errors introduced by order reduction of the subsystems. In fact, it is entirely possible that the resulting interconnected system may not be a good approximation of the original system at all. We will show how combining the $(\gamma,\delta)$-ROM approach with the modular approach solves this problem. 
Specifically, given an interconnected system $\Sigma$ of $N$ subsystems $\Sigma_i^{n_i}$, $i\in\{1,\dots,N\}$, we obtain a modular ROM $\Sigma_{\text{mod}}$ of $\Sigma$  as follows. First, for every subsystem $\Sigma_i^{n_i}$, we determine a $(\gamma_i,\delta_i)$-ROM, denoted as $\Sigma_i^{r_i}$, by solving the optimization problem $\mathcal{P}_1$ defined in Section \ref{sec:MOR}. Then, we connect the obtained $\Sigma_i^{r_i}$ with the same interconnection structure as $\Sigma$ (cf. Figure \ref{fig:modular}). To establish the theoretical bounds on the output of a modular ROM of an interconnected system, we assume the following throughout this section.

\begin{assumption}\label{assum:compositionality_assump}
    Every subsystem $\Sigma_i^{n_i}$, $i\in\{1,\dots,N\}$, is $0$-asymptotically stable. 
    Further, a $(\gamma_i,\delta_i)$-ROM $\Sigma^{r_i}_i$ exists and is known for each subsystem $\Sigma^{n_i}_i$. Finally, for each subsystem $\Sigma_i^{n_i}$, the constants $l_i$ and $k_i$ defined in Lemma \ref{lem:prelim_output_bound} are known.
\end{assumption}

We will first characterize an upper bound on the output of a modular ROM of an arbitrary interconnected system. We will then establish that, for series, parallel, and feedback connections, our approach yields a $(\gamma,\delta)$-modular ROM.

\subsection{Systems with Arbitrary Interconnection Structure}
The following result characterizes an upper bound on the internal outputs of an interconnected system.

\begin{lemma}\label{lem:arbitrary_interconnection}
    Consider a system $\Sigma$ consisting of $N$ subsystems $\Sigma_i^{n_i}, 1\leq i\leq N$ and suppose that Assumption \ref{assum:compositionality_assump} holds.
    Let $\Sigma_{\text{mod}}$ denote an interconnected system, with the same interconnection structure as $\Sigma$ but with subsystems $\Sigma_i^{r_i}$. Further, let $l\coloneqq \sum_{i=1}^N l_i$ where $l_i, \forall i\in\{1,\dots,N\}$ are defined as in Lemma \ref{lem:prelim_output_bound}. Then
    \begin{multline*}
        (1-l)\sum_{i=1}^N\sum_{j=1.j\neq i}^N\norm{z_{ij}}^2+\norm{z_{ij}'}^2 \\
        \leq \sum_{i=1}^N \left( k_i\norm{d_i}^2+l_i\norm{u_i}^2+l_i\norm{u_i'}^2 \right).
    \end{multline*}
\end{lemma}
\begin{proof}
Since each subsystem $1\leq i\leq N$ is $0$-asymptotically stable, it follows from Lemma \ref{lem:prelim_output_bound} that, for each subsystem $i$, there exists constants $l_i$ and $k_i$ such that
\begin{equation}\label{eq:intercon_output_bound}
\begin{split}
    &\sum_{j=1, j\neq i}^N\norm{z_{ij}}^2 + \norm{z_{ij}'}^2 \leq l_i\left(\norm{u_i}^2+\norm{u_i'}^2\right) -\norm{y_i}^2\\
    & -\norm{y_i'}^2+l_i\sum_{j=1, j\neq i}^N\left(\norm{w_{ij}}^2 + \norm{w_{ij}'}^2 \right)+k_1\norm{d_1}^2.
\end{split}
\end{equation}
Adding equation \eqref{eq:intercon_output_bound} for each $i\in\{1,\dots,N\}$ and using the fact that $w_{ij} = z_{ji}$ establishes the claim.
\end{proof}
Let $y_{\text{int}}\coloneqq \text{col}(y_1,\dots,y_N)$ and $y'_{\text{int}}\coloneqq \text{col}(y_1',\dots,y_N')$, where $y_i$ (resp. $y_i'$) denote the external output of subsystem $\Sigma_i^{n_i}$ (resp. $\Sigma_i^{r_i}$). Then the following result provides an error bound between an interconnected system $\Sigma$ with an arbitrary interconnection structure and its ROM $\Sigma_{\text{mod}}$ constructed using the modular approach described above.
\begin{theorem}\label{thm:general_interconnection}
    Consider a system $\Sigma$ consisting of $N$ subsystems $\Sigma_i^{n_i}, 1\leq i\leq N$, and suppose that Assumption \ref{assum:compositionality_assump} holds.
    Let $\Sigma_{\text{mod}}$ denote an interconnected system, with the same interconnection as $\Sigma$ but with subsystems $\Sigma_i^{r_i}$. If $l<1$ and $\gamma_i<1$, for each $i\in\{1,\dots,N\}$, then
    \begin{multline*}
        \norm{y_{\text{int}} - y_{\text{int}}'}^2  \leq 
        \sum_{i=1}^N\left( \left(\mu_i+\frac{k_i\delta_{\text{max}}}{1-l}\right)\norm{d_i}^2 - \eta_i\norm{d_i'}^2 \right)\\+\sum_{i=1}^N\left(\gamma_i\norm{u_i-u_i'}^2 + \left(\delta_i + \frac{l_i\delta_{\text{max}}}{1-l}\right)\norm{\begin{bmatrix}
            u_i \\ u_i'
        \end{bmatrix}}^2\right),
    \end{multline*}
    where $\delta_{\text{max}} = \max\{\delta_1,\dots,\delta_N\}$ and $l$ is defined in Lemma \ref{lem:arbitrary_interconnection}.
\end{theorem}
\begin{proof}
    Since $\Sigma_1^{r_1}$ is a $(\gamma_1,\delta_1)$-ROM of $\Sigma_1^{n_1}$, the following holds:       \begin{multline}\label{eq:arbitrary_interconnection_1}
            \norm{
               \text{col}\left(y_1 - y_1',
               z_{12} - z_{12}',\dots,
               z_{1N} - z_{1N}'\right)
            }^2 \leq (\mu_1-\epsilon_1)\norm{d_1}^2\\
            + \gamma_1\norm{u_1-u_1'}^2+ \gamma_1\sum_{j=2}^N\norm{w_{1j}-w_{1j}'}^2-\eta_1\norm{d_1'}^2\\
             +(\delta_1-\epsilon_1)\norm{\text{col}\left(
                u_1,u_1', w_{12}, w_{12}', \dots,w_{1N}'\right)}^2.
        \end{multline}
        An analogous equation can be obtained for each subsystem $i$ yielding a total of $N$ such equations. Adding these $N$ equations and using the fact that $z_{ij}=w_{ji}$ yields
        \begin{align*}
            &\sum_{i=1}^N\left(\norm{y_i-y_i'}^2+ (1-\gamma_i)\norm{w_{ij}-w_{ij}'}^2 \right) \\
            &\leq \sum_{i=1}^N\gamma_i\norm{u_i-u_i'}^2 +\sum_{i=1}^N\delta_i\left(\norm{u_i}^2+\norm{u_i'}^2\right) \\
            &+\max\{\delta_1,\dots,\delta_N\}\sum_{i=1}^N\sum_{j=1,j\neq i}^N\norm{z_{ij}}^2+\norm{z_{ij}'}^2\\
            &+\sum_{i=1}^N\left(\mu_i\norm{d_i}^2-\eta_i\norm{d_i'}^2 \right).
        \end{align*}
The claim then follows by using Lemma \ref{lem:arbitrary_interconnection} and the fact that $\gamma_i\leq 1$ for all $i\in\{1,\dots,N\}$.
\end{proof}
 
Recall that Theorem \ref{thm:general_interconnection} holds for systems with an arbitrary interconnection structure. Since we do not assume anything about the interconnections, Theorem \ref{thm:general_interconnection} requires that $\gamma_i<1$ and $l_i<1$ for all $i\in\{1,\dots,N\}$. Since each $\gamma_i$ is obtained by solving problem $\mathcal{P}_1$, requiring $\gamma_i<1$ for each subsystem is not restrictive. However, requiring $l<1$ may be restrictive. 
In the next subsection, we will see that by considering specific interconnections such as series, parallel, and feedback, these requirements are either not required or can be relaxed.

\subsection{Systems with Series, Parallel, or Feedback Connections}
\begin{figure}
\begin{subfigure}{0.45\columnwidth}
  \centering
  \includegraphics[width=\linewidth]{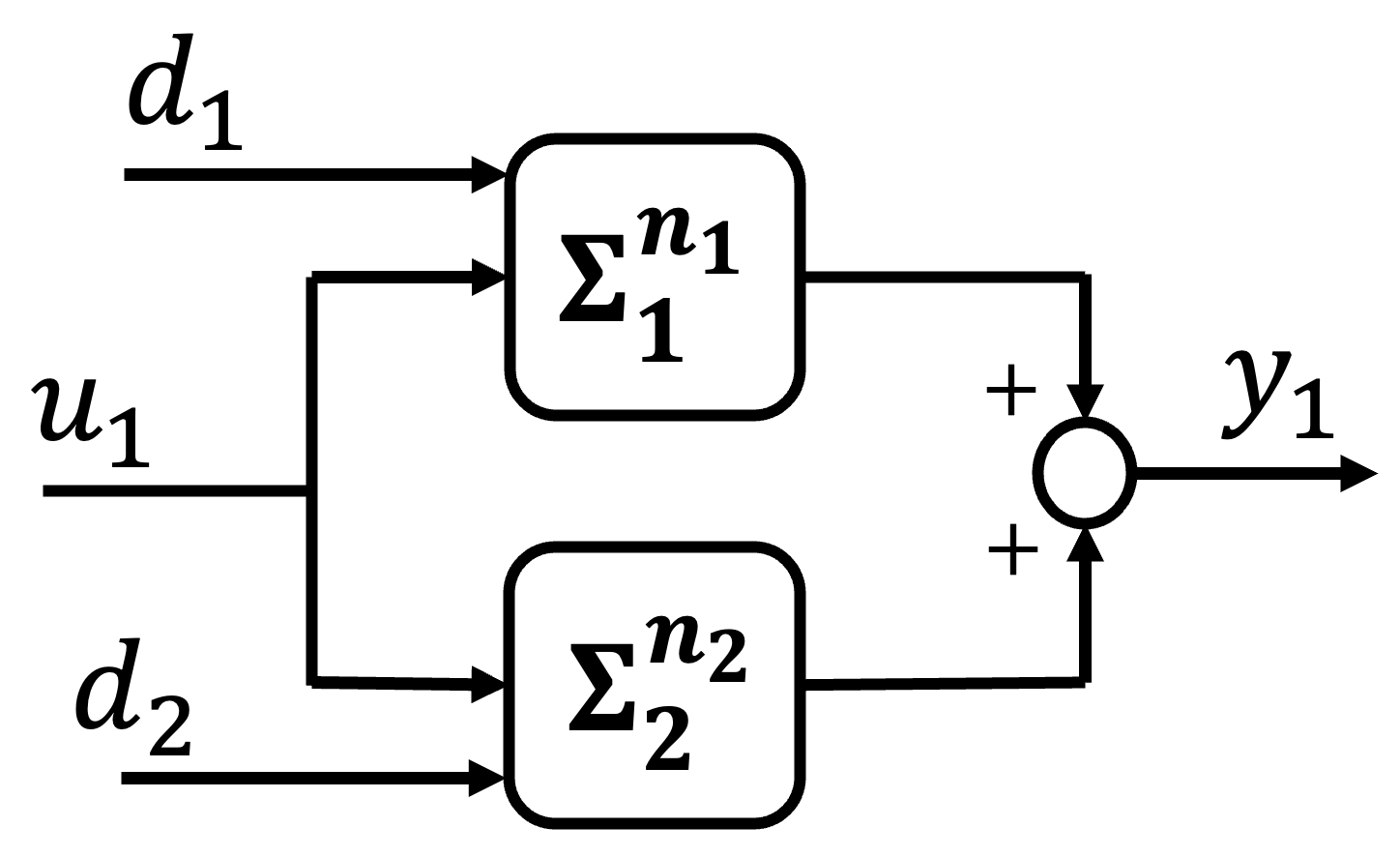}
  \caption{System $\Sigma$.}
  \label{fig:fparallel_FOM}
\end{subfigure}
\hfill
\begin{subfigure}{0.45\columnwidth}
  \centering
  \includegraphics[width=\linewidth]{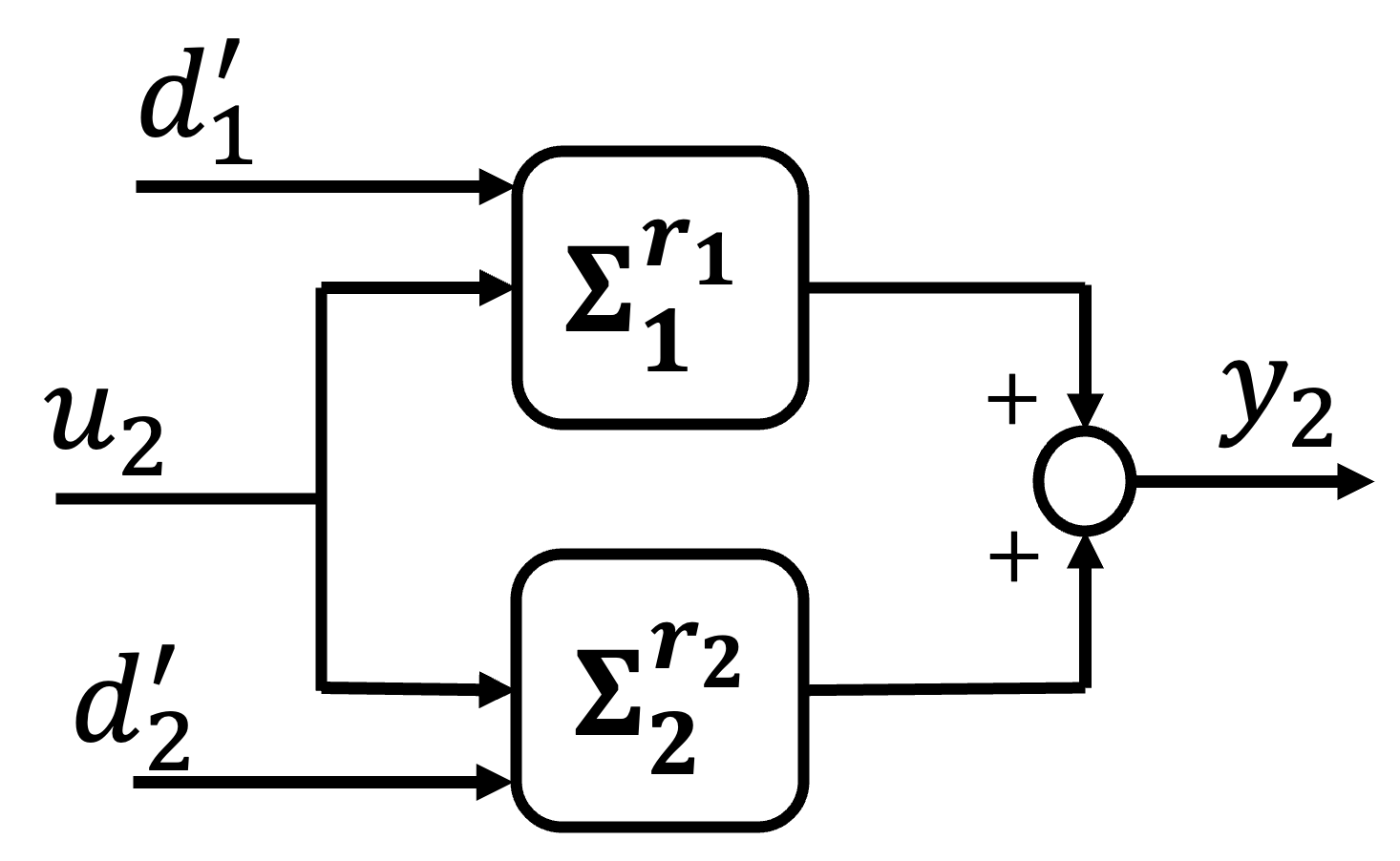}
  \caption{System $\Sigma_{\text{mod}}$.}
  \label{fig:parallel_ROM}
\end{subfigure}
\caption{Parallel interconnection of $N=2$ systems. System $\Sigma$ consists of two FOMs $\Sigma_1^{n_1}$ and $\Sigma_2^{n_2}$ connected in parallel. For $i\in \{1,2\}$, $\Sigma_i^{r_i}$ denotes a $(\gamma_i,\delta_i)$-ROM of $\Sigma_i^{n_i}$. Systems $\Sigma_i^{r_i}$ are connected in parallel to obtain $\Sigma_{\text{mod}}$.}
\label{fig:parallel}
\end{figure}

\begin{figure}[t]
\begin{subfigure}{0.49\columnwidth}
  \centering
  \includegraphics[width=\linewidth]{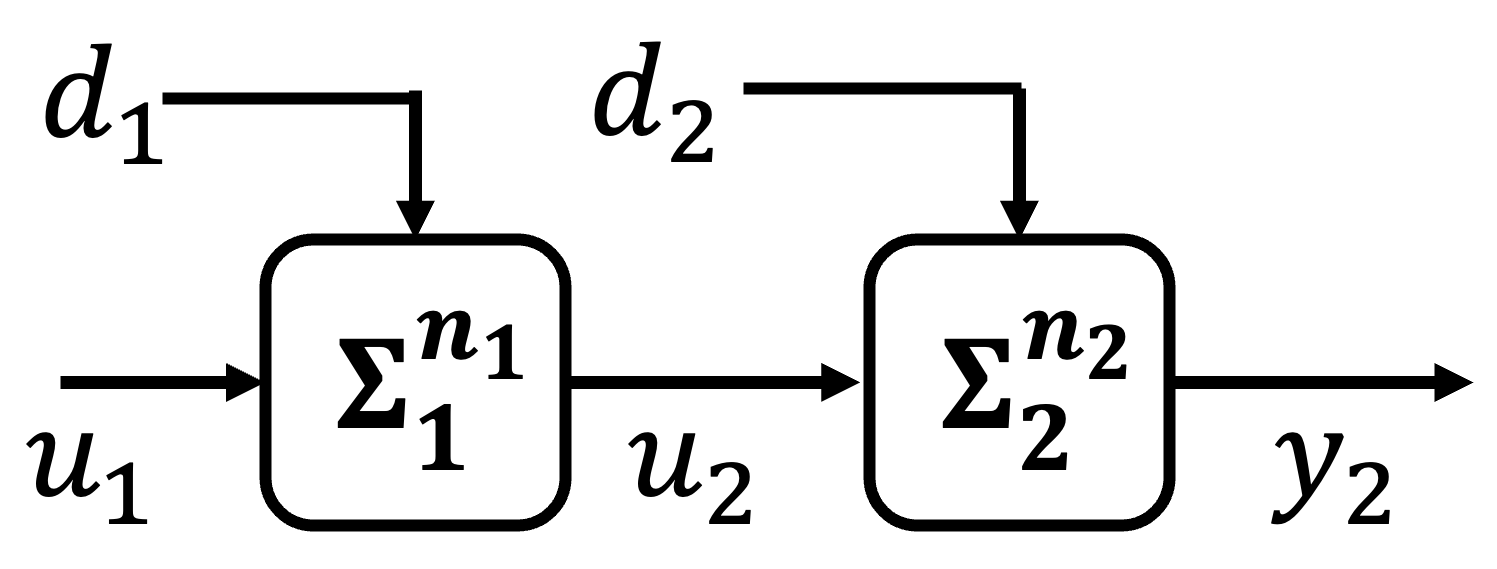}
  \caption{System $\Sigma$.}
  \label{fig:series_FOM}
\end{subfigure}
\hfill
\begin{subfigure}{0.49\columnwidth}
  \centering
  \includegraphics[width=\linewidth]{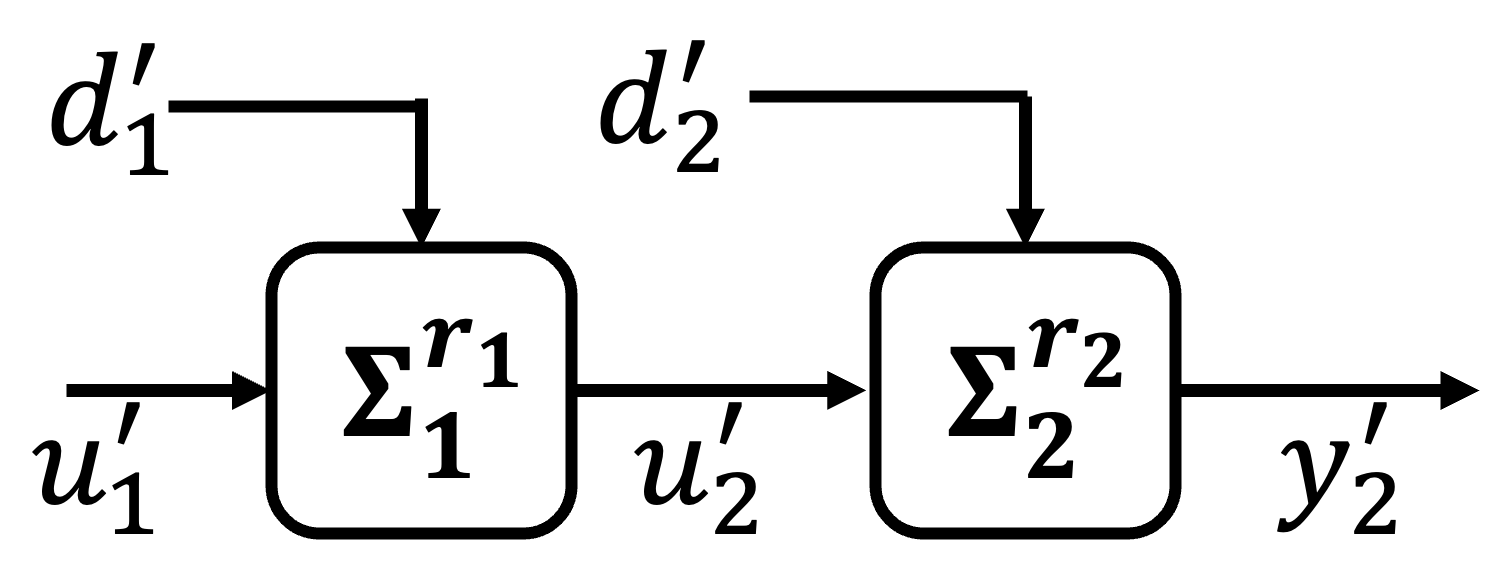}
  \caption{System $\Sigma_{\text{mod}}$.}
  \label{fig:series_ROM}
\end{subfigure}
\caption{Series interconnection of $N=2$ systems. System $\Sigma$ consists of two FOMs $\Sigma_1^{n_1}$ and $\Sigma_2^{n_2}$ connected in series. For $i\in \{1,2\}$, $\Sigma_i^{r_i}$ denotes a $(\gamma_i,\delta_i)$-ROM of $\Sigma_i^{n_i}$. Systems $\Sigma_i^{r_i}$ are connected in series to obtain $\Sigma_{\text{mod}}$.}
\label{fig:series}
\end{figure}

In this subsection, we focus on the most common interconnection structures --  parallel, series, and  feedback -- and establish that the modular ROM $\Sigma_{\text{mod}}$ obtained above is a $(\gamma_r,\delta_r)$-ROM of the interconnected system $\Sigma$, for specified values of $\gamma_r$ and $\delta_r$. We begin with the subsystems being connected in parallel (cf. Figure \ref{fig:parallel}), followed by the subsystems being connected in series (cf. Figure \ref{fig:series}).

\begin{theorem}\label{thm:parallel}
    Consider a system $\Sigma$ consisting of $N$ subsystems $\Sigma_i^{n_i}$, $i\in\{1,\dots,N\}$, connected in parallel and suppose that Assumption \ref{assum:compositionality_assump} holds.  Further, let $\Sigma_{\text{mod}}$ denote a system consisting of $\Sigma_i^{r_i}$ connected in the same parallel interconnection structure as in $\Sigma$.
    Then, for $\gamma_r\coloneqq \max\{\gamma_1,\dots,\gamma_N\}$ and $\delta_r\coloneqq \max\{ \delta_1,\dots,\delta_N\}$, $\Sigma_{\text{mod}}$ is a $(\gamma_r,\delta_r)$-ROM of $\Sigma$.
\end{theorem}
\begin{proof}
The claim follows directly from Theorem \ref{thm:general_interconnection} by substituting $w_{ij}=z_{ji}=0$, $w_{ij}'=z_{ji}'=0$, and, for all $i\in\{1,\dots,N\}$, substituting $u_i=u$ and $u_i'=u'$.
\end{proof}

\begin{theorem}\label{thm:series}
    Consider a system $\Sigma$ consisting of $N$ subsystems $\Sigma_i^{n_i}$, $i\in\{1,\dots,N\}$ connected in series and suppose Assumption \ref{assum:compositionality_assump} holds.
    Further, let $\Sigma_{\text{mod}}$ denote a system consisting of $\Sigma_i^{r_i}$ connected in the same series interconnection as in $\Sigma$.
    Then, for $\gamma_r\coloneqq \prod_{i=1}^N \gamma_i$ and $\delta_r\coloneqq \sum_{i=1}^N\left( \delta_i\prod_{j=i+1}^N\gamma_j\prod_{k=1}^{i-1}l_k\right)$, $\Sigma_{\text{mod}}$ is a $(\gamma_r,\delta_r)$-ROM of $\Sigma$.
\end{theorem}
\begin{proof}
    The claim follows directly from Theorem \ref{thm:general_interconnection} by substituting $w_{ij}=z_{ji}=0$, $w_{ij}'=z_{ji}'=0$, and, for $i\in\{2,\dots,N\}$, substituting $u_i'=y_{i-1}'$ and $u_i=y_{i-1}$.
\end{proof}
\begin{remark}
Theorem \ref{thm:parallel} and Theorem \ref{thm:series} hold for arbitrary values of $\gamma_i>0$ and $l_i>0$, $i\in\{1,\dots,N\}$, i.e., Theorem \ref{thm:parallel} and Theorem \ref{thm:series} do not require $\gamma_i<1$ and $l_i<1$.
\end{remark}

We now consider systems connected in feedback. For $N$ systems connected in negative feedback, i.e., when every system is in a feedback to another system, 
we can obtain the same conditions as in Theorem \ref{thm:general_interconnection} and an analogous error bound. However, relaxed conditions on $\gamma_i$ and $l_i$, $i\in\{1,2\}$, compared to those in Theorem \ref{thm:general_interconnection} were established for $N=2$ systems in \cite[Proposition 6]{pirastehzad2024comparison} which we state below.
\begin{theorem}\label{thm:feedback}
    Consider a system $\Sigma$ consisting of $N=2$ subsystems $\Sigma_i^{n_i}$ connected in feedback and suppose that Assumption \ref{assum:compositionality_assump} holds. Further, let $\Sigma_{\text{mod}}$ denote an interconnected system consisting of $\Sigma_i^{r_i}$ connected in the same feedback interconnection as $\Sigma$. Finally, for a given $\tilde{\epsilon}=1+\nu$, where $\nu>0$ is a constant such that $\tilde{\epsilon}^2\gamma_1\gamma_2<1$ and $\tilde{\epsilon}^2 l_1l_2<1$ holds, $\Sigma_{\text{mod}}$ is a $(\gamma_r,\delta_r)$-ROM of $\Sigma$, where 
    \begin{align*}
        \gamma_r &= \frac{\tilde{\epsilon}\max\{\gamma_1,\gamma_2\}+\tilde{\epsilon}^2\gamma_1\gamma_2}{\nu(1-\gamma_1\gamma_2\tilde{\epsilon}^2)},\\
        \delta_r &= \frac{2\tilde{\epsilon}\max\{\delta_1,\delta_2\}\max\{l_1,l_2\}(1+\tilde{\epsilon}\max\{\gamma_1,\gamma_2\})}{(1-\gamma_1\gamma_2\tilde{\epsilon}^2)(1-l_1l_2\tilde{\epsilon}^2)}\\
        &+ \frac{2\max\{\delta_1,\delta_2\}(1+\tilde{\epsilon}\max\{\gamma_1,\gamma_2\})\left(\nu+\tilde{\epsilon}^2(1-\nu)l_1l_2 \right)}{\nu(1-\gamma_1\gamma_2\tilde{\epsilon}^2)(1-l_1l_2\tilde{\epsilon}^2)}.
    \end{align*}
\end{theorem}
\begin{proof}
    The proof is analogous to that of \cite[Proposition 6]{pirastehzad2024comparison} and has been omitted for brevity.
\end{proof}
Till this point, by restricting ourselves to the case when the FOMs are assumed to be $0$-asymptotically stable, we have shown that the approach proposed in this work satisfies the desirable properties (1)-(4) described in Problem \ref{prob:1}. In the next section, we will establish that the proposed approach also satisfies the desirable property (5) described in Problem \ref{prob:1}.

\section{Closed-Loop Stability of Original System}\label{sec:closed_loop}
In this section, we will establish that a stabilizing controller $\Sigma_K$ designed for a $(\gamma,\delta)$-ROM $\Sigma_r$ can ensure that the output of the FOM $\Sigma^n$ with the same controller is also bounded. Recall that the notion of $(\gamma,\delta)$-similarity requires the systems to be $0$-asymptotically stable. Under this assumption, establishing that the output of the (closed-loop) original system is bounded is not very meaningful. Hence, in this section, we will first generalize the notion of  $(\gamma,\delta)$-similarity of two systems with arbitrary initial conditions and that may not be $0$-asymptotically stable. Specifically, consider two systems $\Sigma_i$, $1\leq i\leq 2$, defined as follows:
\begin{equation}\label{eq:systems}
    \Sigma_i : \begin{cases}
        \dot{x}_i &= A_i x_i + B_iu_i + E_id_i, \quad x_i(0) = x_{i,0}\\
        y_i &= C_ix_i,
    \end{cases}
\end{equation}
where the matrices $A_{i}$ need not be Hurwitz. We first modify Definition \ref{def:similar_systems} to incorporate the non-zero initial conditions. Doing so will require us to both modify the intermediate results from \cite{pirastehzad2024comparison}, as well as establish additional results. 
However, as we will see in Lemma \ref{lem:LMI_unstable},  generalizing the notion of $(\gamma,\delta)$-similarity in this way leads to the same LMI condition as when the systems are assumed to be stable and with zero initial conditions (refer to equation \eqref{eq:LMI_prelim}). Intuitively, this is because the notion of $(\gamma,\delta)$-similarity is inspired from $\mathcal{H}_{\infty}$ control theory, which does not require the assumptions imposed in \cite{pirastehzad2024comparison}. 
We begin with the definition of $(\gamma,\delta)$-similar systems for arbitrary initial conditions and possibly unstable systems. 
\begin{definition}\label{def:similarity_def_unstable}
    Given systems $\Sigma_1^{n_1}$ and $\Sigma_2^{n_2}$, for $\gamma,\delta>0$, the system $\Sigma_2^{n_2}$ is said to be $(\gamma,\delta)$-similar to $\Sigma_1^{n_1}$, if there exist positive constants $\epsilon, \eta, \mu$ and a matrix $K\succ 0$ such that, for every input $u_1, u_2\in \mathcal{L}_2$ and every disturbance $d_1\in \mathcal{L}_2$, there exists a disturbance $d_2\in\mathcal{L}_2$ such that 
    \begin{equation}\label{eq:similarity_def_unstable}
    \begin{split}
        \norm{y_1-y_2}^2 &\leq \gamma\norm{u_1-u_2}^2+(\delta-\epsilon)\norm{\text{col}(u_1,u_2)}\\
        &+(\mu-\epsilon)\norm{d_1}^2-\eta\norm{d_2}^2 + x_s'Kx_s,
    \end{split}
    \end{equation}
    where $x_s\coloneqq \text{col}(x_{1,0},x_{2,0})$.
\end{definition}

Definition \ref{def:similarity_def_unstable} provides a measure on the similarity of general responses as opposed to merely the forced response between two systems. We now provide a characterization of $(\gamma,\delta)$-similarity for systems that may not be $0$-asymptotically stable while deferring the intermediate results to Appendix \ref{sec:appen_generalization}. 
\begin{lemma}\label{lem:LMI_unstable}
    Suppose that the conditions specified in Lemma \ref{lem:LQ_stability} and Theorem \ref{thm:similarity_unstable_riccati} hold. For $\gamma,\delta>0$, $\Sigma_2^{n_2}$ is $(\gamma,\delta)$-similar to $\Sigma_1^{n_1}$ if there exists a positive definite matrix $X$, a matrix $\Pi$, and positive scalars $\eta$ and $\mu$ such that
    \begin{align*}
        &\begin{bmatrix}
        AX + B\Pi + (AX+B\Pi)^\top & * & *\\
        E^\top & -Q(\gamma,\delta) & * \\
        CX+D\Pi & \mathbf{0} & -R
    \end{bmatrix} \prec 0.
    \end{align*}
\end{lemma}
\begin{proof}
    The proof has been omitted for brevity as it is analogous to that of \cite[Theorem 2]{pirastehzad2024comparison}. Note that the intermediate results required for this proof are different than in \cite{pirastehzad2024comparison} and have been provided in Appendix \ref{sec:appen_generalization}. 
\end{proof}

Note that the LMI in Lemma \ref{lem:LMI_unstable} is exactly the same as in Theorem \ref{thm:LMI_prelim}.   We can now utilize the LMI characterized in Lemma \ref{lem:LMI_unstable} to determine a $(\gamma,\delta)$-ROM $\Sigma^r$ of a FOM $\Sigma^{n}$. 
\begin{theorem}
    Given a system $\Sigma^n$ and a positive integer $r<n$, suppose that a solution $\{A_2^*, B_2^*,C_2^*, X^*, \Pi^*, \gamma^*, \eta^*\}$ to problem $\mathcal{P}_1$ exists. Then, the obtained ROM $\Sigma^r$ defined by the system matrices $A_2^*, B_2^*, C_2^*$ is a $(\gamma^*,\delta^*)$-ROM of $\Sigma^n$.
\end{theorem}
\begin{proof}
    The proof is analogous to that of Theorem \ref{thm:generalized_ROM_stable} and has been omitted for brevity.
\end{proof}

Having extended the notion of $(\gamma,\delta)$-similarity to unstable systems, we can now characterize the closed-loop stability of the original system when a controller, denoted as $\Sigma_K$, designed for the $(\gamma,\delta)$-ROM $\Sigma^r$ is used for the original system.  
Note that, since we do not impose any stability assumption on the FOM $\Sigma^n$, it is possible that the obtained ROM $\Sigma^r$ may also be unstable. If $\Sigma^r$ or $\Sigma^n$ is not stabilizable, then designing a controller $\Sigma^K$ is not meaningful. Hence, a necessary assumption is that the FOM $\Sigma^n$ and its obtained ROM $\Sigma^r$ are stabilizable.
Finally, since in this work we restrict ourselves to the class of LTI systems, we  consider only linear controllers $\Sigma_K$. We state this assumption more formally below.

\begin{assumption}\label{assum:controller}
Given a system $\Sigma^n$ of order $n$, there exists a stabilizable $(\gamma_1,\delta_1)$-ROM $\Sigma^r$ of order $r<n$. 
Thus, a (stabilizing) controller $\Sigma_K$ exists for system $\Sigma^r$ and is $0$-asymptotically stable. 
\end{assumption}

Given the assumption that $\Sigma_K$ is $0$-asymptotically stable (see Assumption \ref{assum:controller}), it follows from \cite[Section 4.1]{desoer2009feedback} that there exist positive constants $l_1$ and $k_1$ such that for all $u_1, d_1 \in \mathcal{L}_2$  \begin{align}\label{eq:controller_output_bound}
        \norm{y_K}^2 \leq l_1\norm{u_K}^2 + k_1\norm{d_K}^2,
    \end{align}
where $y_K$, $u_K$, and $d_K$ denotes the output, input, and disturbance vectors of $\Sigma_K$. Further, since the same controller is applied to the both $\Sigma^r$ and $\Sigma^n$, from Lemma \ref{lem:prelim_system_iteslef}, there exists a $\gamma_K$ such that for any $\delta_K>0$, the controller $\Sigma_K$ is $(\gamma_K,\delta_K)$-similar to itself.

    

\begin{figure}
\begin{subfigure}{0.45\columnwidth}
  \centering
  \includegraphics[width=\linewidth]{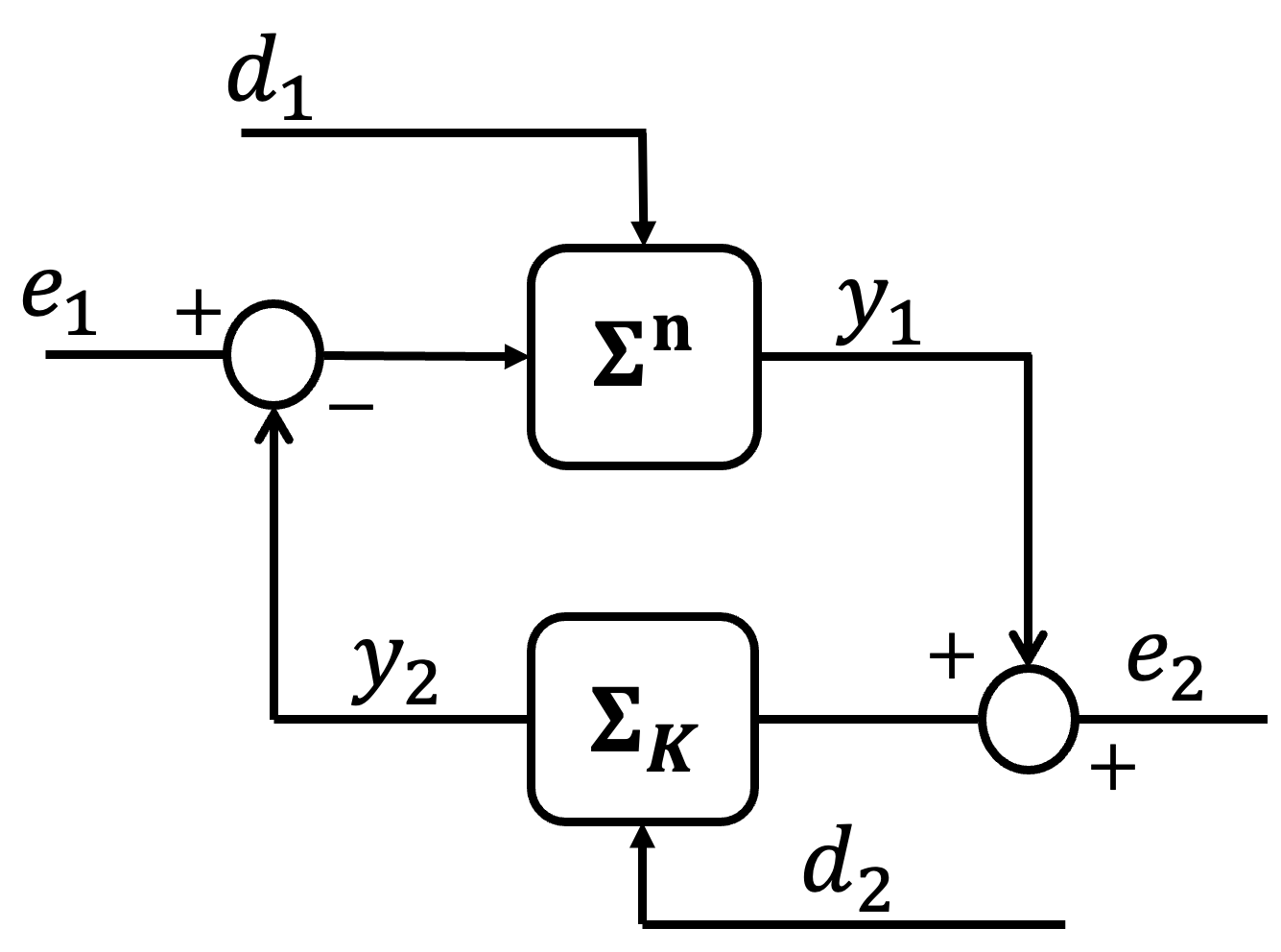}
  \caption{Feedback interconnection of $\Sigma^r$ and $\Sigma_K$.}
  \label{fig:fb_reduced}
\end{subfigure}
\hfill
\begin{subfigure}{0.49\columnwidth}
  \centering
  \includegraphics[width=\linewidth]{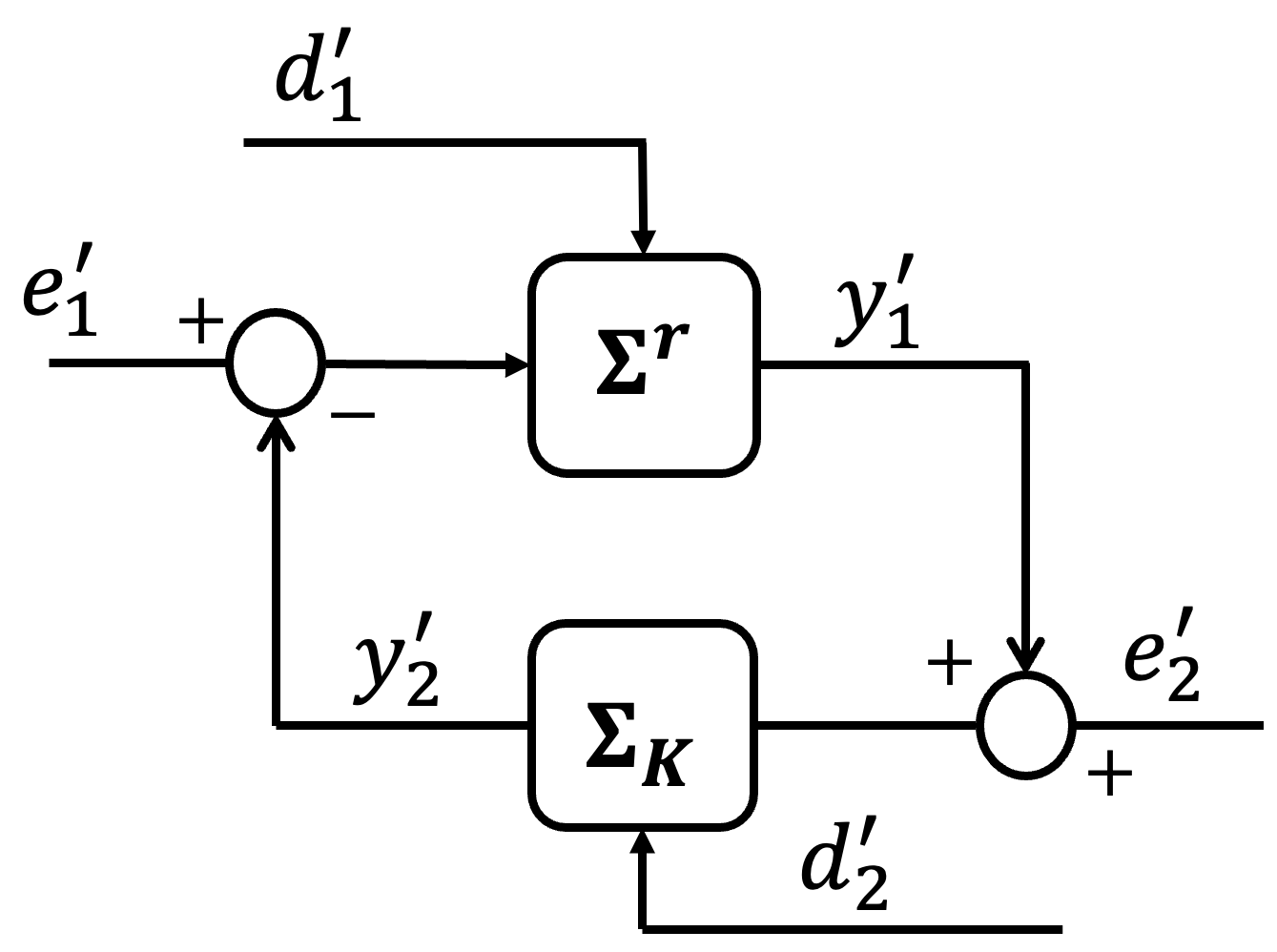}
  \caption{Feedback interconnection of $\Sigma^n$ and $\Sigma_K$.}
  \label{fig:fb_original}
\end{subfigure}
\caption{Controller $\Sigma_K$ designed for ROM $\Sigma^r$ connected in feedback with ROM $\Sigma^r$ and original system $\Sigma^n$.}
\label{fig:closed_loop_stability}
\end{figure}
For $l_1$ and $k_1$ defined in \eqref{eq:controller_output_bound}, define $$\hat{Q} := \begin{bmatrix}
    2(\gamma_K+\delta_K+l_1)\mathbf{I} & -2\gamma_K\mathbf{I}\\
    -2\gamma_K\mathbf{I} & 2(\gamma_K+\delta_K+l_1)\mathbf{I}
\end{bmatrix},$$ and let $\tilde{\epsilon}\coloneqq 1+\nu$ for some constant $\nu>0$. Further let $l\coloneqq l_1 + \lambda_{\text{max}}(\hat{Q})$ and $k\coloneqq k_1 + 2(\mu+k_1)$. 
We are now ready to establish the main result of this section, i.e., a controller designed to stabilize the $(\gamma,\delta)$-ROM $\Sigma^r$ when applied to the FOM $\Sigma^n$ (cf. Figure \ref{fig:closed_loop_stability}), ensures that the output of $\Sigma^n$ is bounded.

\begin{theorem}\label{thm:closed_loop_stability}
    Given a FOM $\Sigma^n$ and a positive integer $r<n$, suppose Assumption \ref{assum:controller} holds. Consider that $\Sigma_K$ is connected in feedback to $\Sigma^n$.
    For some constant $\nu>0$, let the following conditions hold: 
    \begin{align}
        1-\tilde{\epsilon}\gamma_1\gamma_K&>0 \label{eq:thm5.2}\\
        1-\sqrt{\frac{2\tilde{\epsilon}\gamma_1(\delta_2-\epsilon_2)+2l(\delta_1-\epsilon_1)}{1-\tilde{\epsilon}^2\gamma_1\gamma_2}}&>0.
    \end{align} 
    Then, the following inequality holds
    \begin{align*}
        &\left(1-\sqrt{\frac{2\tilde{\epsilon}\gamma_1(\delta_2-\epsilon_2)+2l(\delta_1-\epsilon_1)}{1-\tilde{\epsilon}^2\gamma_1\gamma_2}} \right)\norm{y_1} \leq \\
        &\sqrt{\frac{\tilde{\epsilon}\gamma_1}{\nu(1-\tilde{\epsilon}^2\gamma_1\gamma_2)}}\norm{e_1-e_1'} + \sqrt{\frac{\tilde{\epsilon}\gamma_1\gamma_2}{\nu(1-\tilde{\epsilon}^2\gamma_1\gamma_2)}}\norm{e_2-e_2'}\\
        & + \sqrt{\frac{2(\delta_1-\epsilon_1)}{1-\tilde{\epsilon}^2\gamma_1\gamma_2}}\norm{\begin{bmatrix}
             e_1 \\ e_1'
         \end{bmatrix}} \\
       & + \sqrt{\left(\frac{2\tilde{\epsilon}\gamma_1(\delta_2-\epsilon_2)+2l(\delta_1-\epsilon_1)}{1-\tilde{\epsilon}^2\gamma_1\gamma_2}\right)}\norm{\begin{bmatrix}
             e_2 \\ e_2'
         \end{bmatrix}}\\
         &+\left( 1+ \sqrt{\frac{2\tilde{\epsilon}\gamma_1(\delta_2-\epsilon_2)+2l(\delta_1-\epsilon_1)}{1-\tilde{\epsilon}^2\gamma_1\gamma_2}}\right)\norm{y_1'}\\
         & + \sqrt{\left(\frac{\mu_1-\epsilon_1+2k(\delta_1-\epsilon_1)}{1-\tilde{\epsilon}^2\gamma_1\gamma_2}\right)}\norm{d_1} +\sqrt{\frac{\eta_1}{1-\tilde{\epsilon}^2\gamma_1\gamma_2}}\norm{d_1'} \\
         & + \sqrt{\frac{\tilde{\epsilon}\gamma_1(\mu_2-\epsilon_2)}{1-\tilde{\epsilon}^2\gamma_1\gamma_2}\norm{d_2}^2  + \frac{\tilde{\epsilon}\gamma_1\eta_2}{1-\tilde{\epsilon}^2\gamma_1\gamma_2}\norm{d_2'}^2}.
    \end{align*}
\end{theorem} 
\begin{proof}
    Since $\Sigma^r$ is $(\gamma_1,\delta_1)$-similar to $\Sigma^n$, from \eqref{eq:similarity_def_unstable} and selecting $u_1=e_1-y_2$ and $u_1' = e_1'-y_2'$ (cf. Figure \ref{fig:closed_loop_stability}) yields
    \begin{align*}
        \norm{y_1-y_1'}^2 &\leq \gamma_1\norm{e_1-y_2-e_1'-y_2'}^2 \\
        &+ (\delta_1-\epsilon_1)
        \norm{\text{col}(e_1-y_2,e_1'-y_2')}\\
        &+ (\mu_1-\epsilon_1)\norm{d_1}^2-\eta_1\norm{d_1'}^2 + x_sKx_s.
    \end{align*}
    Using the triangle and Cauchy-Schwarz inequalities yield
    \begin{equation}\label{eq:output_y1}
    \begin{split}
        \norm{y_1-y_1'}^2 & \leq \frac{\tilde{\epsilon}}{\nu}\gamma_1\norm{e_1-e_1'}^2 + 2(\delta_1-\epsilon_1)\norm{\text{col}(e_1,e_1')}\\
        & +\tilde{\epsilon}\gamma_1\norm{y_2-y_2'}^2 + 2(\delta_1-\epsilon_1)\norm{\text{col}(y_2,y_2')}\\
        & + (\mu_1-\epsilon_1)\norm{d_1}^2-\eta_1\norm{d_1'}^2.
    \end{split}
    \end{equation}
    Analogously, we obtain
    \begin{equation}\label{eq:output_y2}
    \begin{split}
        \norm{y_2-y_2'}^2 & \leq \frac{\tilde{\epsilon}}{\nu}\gamma_2\norm{e_2-e_2'}^2 + 2(\delta_2-\epsilon_2)\norm{\text{col}(e_2,e_2')}\\
        & +\tilde{\epsilon}\gamma_2\norm{y_1-y_1'}^2 + 2(\delta_2-\epsilon_2)\norm{\text{col}(y_1,y_1')}\\
        & + (\mu_2-\epsilon_2)\norm{d_2}^2-\eta_2\norm{d_2'}^2.
    \end{split}
    \end{equation}
    Substituting \eqref{eq:output_y2} in \eqref{eq:output_y1} and rearranging terms yields
    \begin{equation}\label{eq:bound_y1_main}
    \begin{split}
         &\norm{y_1-y_1'}^2 \leq -\frac{\eta_1}{1-\tilde{\epsilon}^2\gamma_1\gamma_2}\norm{d_1'}^2 - \frac{\tilde{\epsilon}\gamma_1\eta_2}{1-\tilde{\epsilon}^2\gamma_1\gamma_2}\norm{d_2'}^2 + \\
         & \frac{\tilde{\epsilon}\gamma_1}{\nu(1-\tilde{\epsilon}^2\gamma_1\gamma_2)}\norm{e_1-e_1'}^2 + \frac{\tilde{\epsilon}\gamma_1\gamma_2}{\nu(1-\tilde{\epsilon}^2\gamma_1\gamma_2)}\norm{e_2-e_2'}^2\\
         & + \frac{2(\delta_1-\epsilon_1)}{1-\tilde{\epsilon}^2\gamma_1\gamma_2}\norm{\begin{bmatrix}
             e_1 \\ e_1'
         \end{bmatrix}}^2 + \frac{2\tilde{\epsilon}\gamma_1(\delta_2-\epsilon_2)}{1-\tilde{\epsilon}^2\gamma_1\gamma_2}\norm{\begin{bmatrix}
             e_2 \\ e_2'
         \end{bmatrix}}^2\\
         & + \frac{2(\delta_1-\epsilon_1)}{1-\tilde{\epsilon}^2\gamma_1\gamma_2}\norm{\begin{bmatrix}
             y_2 \\ y_2'
         \end{bmatrix}}^2 + \frac{2\tilde{\epsilon}\gamma_1(\delta_2-\epsilon_2)}{1-\tilde{\epsilon}^2\gamma_1\gamma_2}\norm{\begin{bmatrix}
             y_1 \\ y_1'
         \end{bmatrix}}^2\\
         & + \frac{\mu_1-\epsilon_1}{1-\tilde{\epsilon}^2\gamma_1\gamma_2}\norm{d_1}^2 + \frac{\tilde{\epsilon}\gamma_1(\mu_2-\epsilon_2)}{1-\tilde{\epsilon}^2\gamma_1\gamma_2}\norm{d_2}^2.
         \end{split}
    \end{equation}
    Under Assumption \ref{assum:controller} and from Lemmas \ref{lem:prelim_system_iteslef} and \ref{lem:prelim_output_bound}, there exists constants $l\coloneqq l_1 + \lambda_{\text{max}}(\hat{Q})$ and $k\coloneqq k_1+2(\mu+k_1)$ such that 
    \begin{align}\label{eq:bound_y1_stacked}
        &\norm{
            \text{col}(y_2, y_2')
        }^2 \leq l\norm{\text{col}(
            e_2-y_1, e_2'-y_1')
        }^2 + k\norm{d_1}^2 \nonumber \\
        & \leq l\norm{
            \text{col}(e_2, e_2')
        }^2 +l\norm{\text{col}(
            y_1, y_1')
        }^2 + k\norm{d_1}^2,
    \end{align}
    where the last inequality is obtained by using triangle inequality.
    Substituting \eqref{eq:bound_y1_stacked} in \eqref{eq:bound_y1_main} and using \eqref{eq:thm5.2}  yields
    \begin{align*}
        &\norm{y_1-y_1'}^2 \leq \frac{\tilde{\epsilon}\gamma_1(\mu_2-\epsilon_2)}{1-\tilde{\epsilon}^2\gamma_1\gamma_2}\norm{d_2}^2  + \frac{\tilde{\epsilon}\gamma_1\eta_2}{1-\tilde{\epsilon}^2\gamma_1\gamma_2}\norm{d_2'}^2 +\\
        & \frac{\tilde{\epsilon}\gamma_1}{\nu(1-\tilde{\epsilon}^2\gamma_1\gamma_2)}\norm{e_1-e_1'}^2 + \frac{\tilde{\epsilon}\gamma_1\gamma_2}{\nu(1-\tilde{\epsilon}^2\gamma_1\gamma_2)}\norm{e_2-e_2'}^2\\
        & + \frac{2(\delta_1-\epsilon_1)}{1-\tilde{\epsilon}^2\gamma_1\gamma_2}\norm{\text{col}(e_1,e_1')}^2\\
         &+\left(\frac{2\tilde{\epsilon}\gamma_1(\delta_2-\epsilon_2)}{1-\tilde{\epsilon}^2\gamma_1\gamma_2}+\frac{2l(\delta_1-\epsilon_1)}{1-\tilde{\epsilon}^2\gamma_1\gamma_2}\right)\norm{\text{col}(e_1,e_2')}^2\\
         &+ \left(\frac{2\tilde{\epsilon}\gamma_1(\delta_2-\epsilon_2)}{1-\tilde{\epsilon}^2\gamma_1\gamma_2} + \frac{2l(\delta_1-\epsilon_1)}{1-\tilde{\epsilon}^2\gamma_1\gamma_2}\right)\norm{\text{col}(y_1,y_1')}^2\\
         & + \left(\frac{\mu_1-\epsilon_1}{1-\tilde{\epsilon}^2\gamma_1\gamma_2}+\frac{2k(\delta_1-\epsilon_1)}{1-\tilde{\epsilon}^2\gamma_1\gamma_2}\right)\norm{d_1}^2 +\frac{\eta_1}{1-\tilde{\epsilon}^2\gamma_1\gamma_2}\norm{d_1'}^2.
    \end{align*}
    Taking square root on both sides and using triangle inequality and the fact that the square root of sum of positive numbers is at most sum of square roots of positive numbers yields
    \begin{align*}
        &\norm{y_1}-\norm{y_1'} \leq \sqrt{\frac{2(\delta_1-\epsilon_1)}{1-\tilde{\epsilon}^2\gamma_1\gamma_2}}\norm{\text{col}(e_1,e_1')}+\\
        &\sqrt{\frac{\tilde{\epsilon}\gamma_1}{\nu(1-\tilde{\epsilon}^2\gamma_1\gamma_2)}}\norm{e_1-e_1'} + \sqrt{\frac{\tilde{\epsilon}\gamma_1\gamma_2}{\nu(1-\tilde{\epsilon}^2\gamma_1\gamma_2)}}\norm{e_2-e_2'}\\
       & + \sqrt{\left(\frac{2\tilde{\epsilon}\gamma_1(\delta_2-\epsilon_2)+2l(\delta_1-\epsilon_1)}{1-\tilde{\epsilon}^2\gamma_1\gamma_2}\right)}\norm{\text{col}(e_2,e_2')}\\
         &+ \sqrt{\left(\frac{2\tilde{\epsilon}\gamma_1(\delta_2-\epsilon_2)+2l(\delta_1-\epsilon_1)}{1-\tilde{\epsilon}^2\gamma_1\gamma_2}\right)}\left(\norm{y_1} + \norm{y_1'}\right)\\
         & + \sqrt{\left(\frac{\mu_1-\epsilon_1+2k(\delta_1-\epsilon_1)}{1-\tilde{\epsilon}^2\gamma_1\gamma_2}\right)}\norm{d_1} +\sqrt{\frac{\eta_1}{1-\tilde{\epsilon}^2\gamma_1\gamma_2}}\norm{d_1'} \\
         & + \sqrt{\frac{\tilde{\epsilon}\gamma_1(\mu_2-\epsilon_2)}{1-\tilde{\epsilon}^2\gamma_1\gamma_2}\norm{d_2}^2  + \frac{\tilde{\epsilon}\gamma_1\eta_2}{1-\tilde{\epsilon}^2\gamma_1\gamma_2}\norm{d_2'}^2}.
    \end{align*}
    Since $\Sigma_K$ is stabilizing for the ROM $\Sigma^r$, it follows that $\norm{y_1'}$ is bounded. Thus, rearranging the terms yields the result.
\end{proof}

\begin{remark} 
    So far in this work we have established that the proposed approach, i.e., the $(\gamma,\delta)$-ROM obtained as a solution to the optimization problem $\mathcal{P}_1$ has the following properties: (1) it minimizes the error between the $l_2$-norm of the outputs of the FOM and the obtained $(\gamma,\delta)$-ROM (cf. Theorem \ref{thm:generalized_ROM_stable}), (2) if the FOM is $0$-asymptotically stable, then the obtained $(\gamma,\delta)$-ROM is $0$-asymptotically stable (cf. Theorem \ref{thm:stability_preserved}), (3) the obtained $(\gamma,\delta)$-ROM is robust to any modeling errors and disturbances $d_1\in\mathcal{L}_2$ (cf. Theorem \ref{thm:generalized_ROM_stable}), (4) facilitates an approach for model reduction of interconnected systems with theoretical guarantees on the error (cf. Theorem \ref{thm:general_interconnection}), and (5) allows the controller designed for the $(\gamma,\delta)$-ROM to be applied to the respective FOM (cf. Theorem \ref{thm:closed_loop_stability}). 
\end{remark}


It might be possible that some of the properties (such as preserving stability) can be achieved by imposing them as a constraint in the optimization problems for the existing methods. However, doing so for all the properties defined in Problem \ref{prob:1} makes the optimization problem  challenging (if not infeasible) to solve. We now show that our proposed approach can also be used in conjunction with existing approaches such as balanced truncation or moment matching. 
This is beneficial because some of the theoretical guarantees established for the proposed approach will now also hold for these methods, while inheriting the desired features of these methods. 

\section{Combining $(\gamma,\delta)$-ROM with existing approaches}\label{sec:Add_on}
In this section, we will modify the optimization problem $\mathcal{P}_1$ 
such that it can be combined with moment matching and balanced truncation. We begin with the former. 

For moment matching, we combine our framework with the approach described in \cite{necoara2019parameter}.
Recall from Section \ref{sec:prelim} that the reduced order system matrices $A_2$ and $C_2$ in moment matching are fully characterized by the matrix $B_2$.
Given an observable pair $(L,S)$ such that $\text{spec}(S)\cap \text{spec}(A_1)=\emptyset$, combining the problem $\mathcal{P}_{1}$ with moment matching yields the following optimization problem, that we refer to as $\mathcal{P}_{\text{MM}}$: 
\begin{align}
    &\textrm{Problem $\mathcal{P}_{\text{MM}}$: }\min_{B_2,X\succ 0,\Pi,\gamma>0,\delta>0} \left(\gamma+\delta\right)\\
    &\text{subject to} \nonumber\\
    & \begin{bmatrix}
        AX + B\Pi + (AX+B\Pi)^\top & * & *\\
        E^\top & -Q(\gamma,\delta) & * \\
        CX+D\Pi & \mathbf{0} & -R
    \end{bmatrix} \prec 0 ,\nonumber
\end{align}
where matrices $A, B, C, D$ and $E$ are as defined in \eqref{eq:composite_matrices_prelim}, matrices $Q$ and $R$ are defined in \eqref{eq:Q_and_R_prelim}, and matrices $A_2$ and $C_2$ are defined in \eqref{eq:family_MM}. Note that, in moment matching, there is an additional constraint that $\text{spec}(S)\cap \text{spec}(S-B_2L)=\emptyset$. Analogous to \cite{necoara2019parameter}, one can select $S$ such that $\text{spec}(S)\subseteq \mathbb{C}^+$. From Theorem \ref{thm:stability_preserved}, since the obtained ROM is guaranteed to be stable, $\text{spec}(S)\cap \text{spec}(S-B_2L)=\emptyset$ will be satisfied. 

\begin{theorem}\label{thm:combined_MM}
    Given a stable system $\Sigma^n$, an observable pair $(S,L)$, and a positive integer $r<n$, suppose that a solution $\{B_2^*,X^*,\Pi^*,\gamma^*,\delta^*\}$ to problem $\mathcal{P}_{\text{MM}}$ exists. Then, the system $\Sigma^r$ defined by the system matrices $B_2^*$ is a $(\gamma^*,\delta^*)$-ROM of $\Sigma^n$ that satisfies properties (1)-(4) described in Problem \ref{prob:1}.
\end{theorem}
\begin{proof}
    Since the obtained ROM is a $(\gamma,\delta)$-ROM, the claim holds.
\end{proof}

\begin{remark}
From Theorem \ref{thm:combined_MM}, combining the proposed framework with moment matching not only provides an error bound, but also ensures that the obtained ROM is stable and robust. To the best of our knowledge, this is the first result that satisfies these properties for moment matching.
\end{remark}

We now describe how the proposed approach can be used with balanced truncation. 
Although balanced truncation preserves stability and provides a theoretical error bound, 
as we will see shortly, combining balanced truncation with the proposed approach is still beneficial as it ensures that the ROM obtained by balanced truncation is robust. 

Since balanced truncation already provides a ROM that satisfies properties (1) and (2), the idea is to utilize this ROM and determine a driving input $d_2\in\mathcal{L}_2$ such that the ROM is $(\gamma,\delta)$-similar to its respective FOM. To this end, consider the following optimization problem that we will refer to as $\mathcal{P}_{\text{BT}}$:
\begin{align}
    &\textrm{Problem $\mathcal{P}_{\text{BT}}$: }\min_{X\succ 0,\Pi,\gamma>0,\delta>0} \left(\gamma+\delta\right)\\
    &\text{subject to} \nonumber\\
    & \begin{bmatrix}
        AX + B\Pi + (AX+B\Pi)^\top & * & *\\
        E^\top & -Q(\gamma,\delta) & * \\
        CX+D\Pi & \mathbf{0} & -R
    \end{bmatrix} \prec 0 ,\nonumber
\end{align}
where matrices $A, B, C, D$ and $E$ are as defined in \eqref{eq:composite_matrices_prelim}, matrices $Q$ and $R$ are defined in \eqref{eq:Q_and_R_prelim}, and matrices $A_2, B_2,$ and $C_2$ are obtained from balanced truncation method. 

\begin{theorem}\label{thm:combined_BT}
    Given a stable system $\Sigma^n$ and its ROM $\Sigma^r$ obtained via balanced truncation, suppose that a solution $\{X^*,\Pi^*,\gamma^*,\delta^*\}$ to the optimization problem $\mathcal{P}_{\text{BT}}$ exist. Then, system $\Sigma^r$ is $(\gamma^*,\delta^*)$-similar to $\Sigma^n$.
    Further, $\Sigma^r$ satisfies properties (1)-(4) described in Problem \ref{prob:1}.
\end{theorem}
\begin{proof}
    Since the given ROM is a $(\gamma,\delta)$-similar to its FOM, the claim holds.
\end{proof}
Note that, to combine moment-matching or balanced truncation with our framework, we do not consider the system matrix $E_1$ associated with the disturbance in the FOM when constructing the ROM. This is because the applying $d_2$ to the ROM accounts for $d_1$. 
\begin{remark}
    Theorem \ref{thm:combined_BT} provides an additional bound on the difference between the outputs of $\Sigma^n$ and $\Sigma^r$. A preexisting bound on the error for the balanced truncation is given in \eqref{eq:bound_BT}. The bound which is lower among the two can be used. 
\end{remark}

\begin{remark}
    The constraint in the optimization problem $\mathcal{P}_{\text{BT}}$ is an LMI constraint for which numerous computational techniques are available.
\end{remark}

\section{Numerical Results}\label{sec:numerics}
We now numerically illustrate the properties established in this work on (i) a cart with double-pendulum model, (ii) a coupled spring-mass-damper system, and (iii) a building model.
For all of the numerical studies in this work, we use MATLAB R2023b (YALMIP \cite{Lofberg2004}, SDPT3 \cite{tutuncu2003solving}, and Gurobi \cite{gurobi}) to solve the specified optimization problem. 

\subsection{Double-pendulum Model}
In this subsection, we will validate properties (1)-(3) by illustrating our approach on a cart with double-pendulum model\cite{6760760}.
The model has order $n=6$ and is defined by the following system matrices:
\begin{equation}\label{eq:dynamics_cart}
\begin{split}
    &A_1 = \begin{bmatrix}
        0 & 1 & 0 & 0 & 0 & 0\\
        -1 & -1 & 98/5 & 1 & 0 & 0\\
        0 & 0 & 0 & 1 & 0 & 0\\
        1 & 1 & -196/5 & -2 & 49/5 & 1\\
        0 & 0 & 0 & 0 & 0 & 1\\
        0 & 0 & 98/5 & 1 & -98/5 & -2
    \end{bmatrix}, \\
    &B_1 = \begin{bmatrix}
        0 & 1 & 0 & -1& 0&0
    \end{bmatrix}^{\top}, \\
    &C_1 = \begin{bmatrix}
        1 & 0 & 0 & 0 & 0 & 0
        \end{bmatrix}, E_1 = C_1^{\top}.
\end{split}
\end{equation}

\begin{figure}
\begin{subfigure}{0.49\columnwidth}
  \centering
  \includegraphics[width=\linewidth]{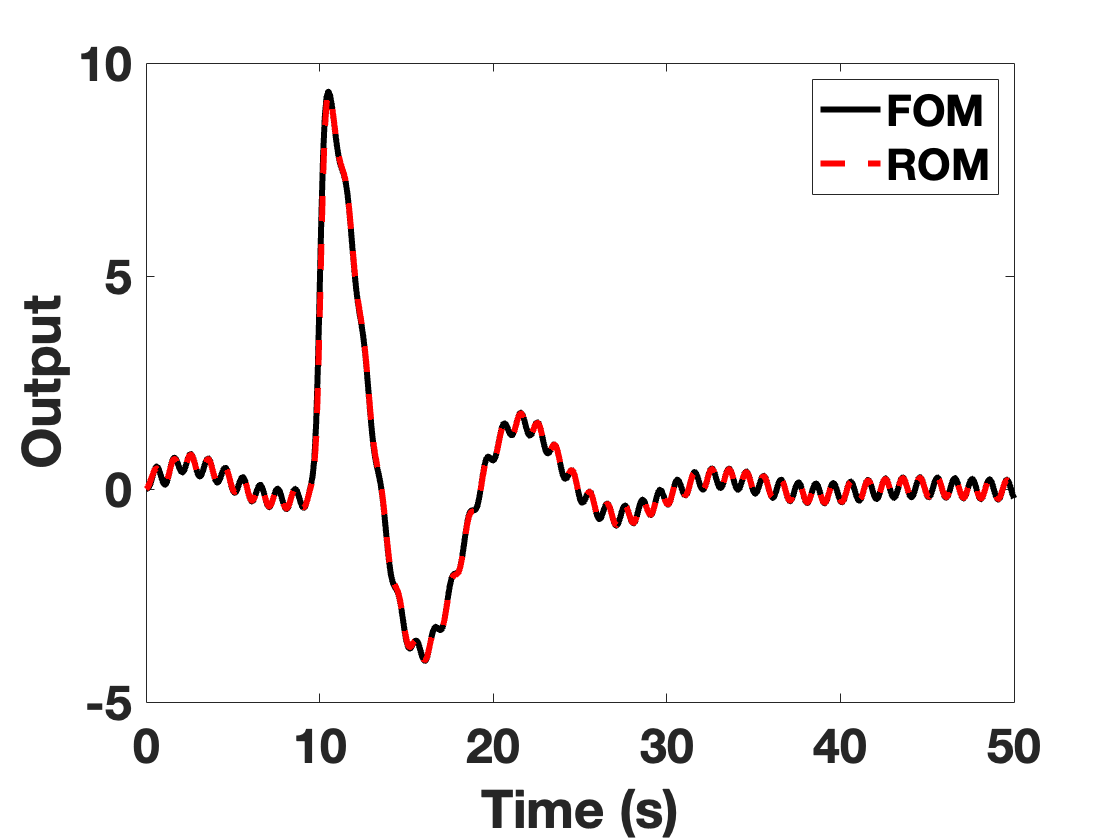}
  \caption{Comparison of outputs of $\Sigma^2_{\text{cart}}$ and $\Sigma^6_{\text{cart}}$.}
  \label{fig:out_cart_gen}
\end{subfigure}
\hfill
\begin{subfigure}{0.49\columnwidth}
  \centering
  \includegraphics[width=\linewidth]{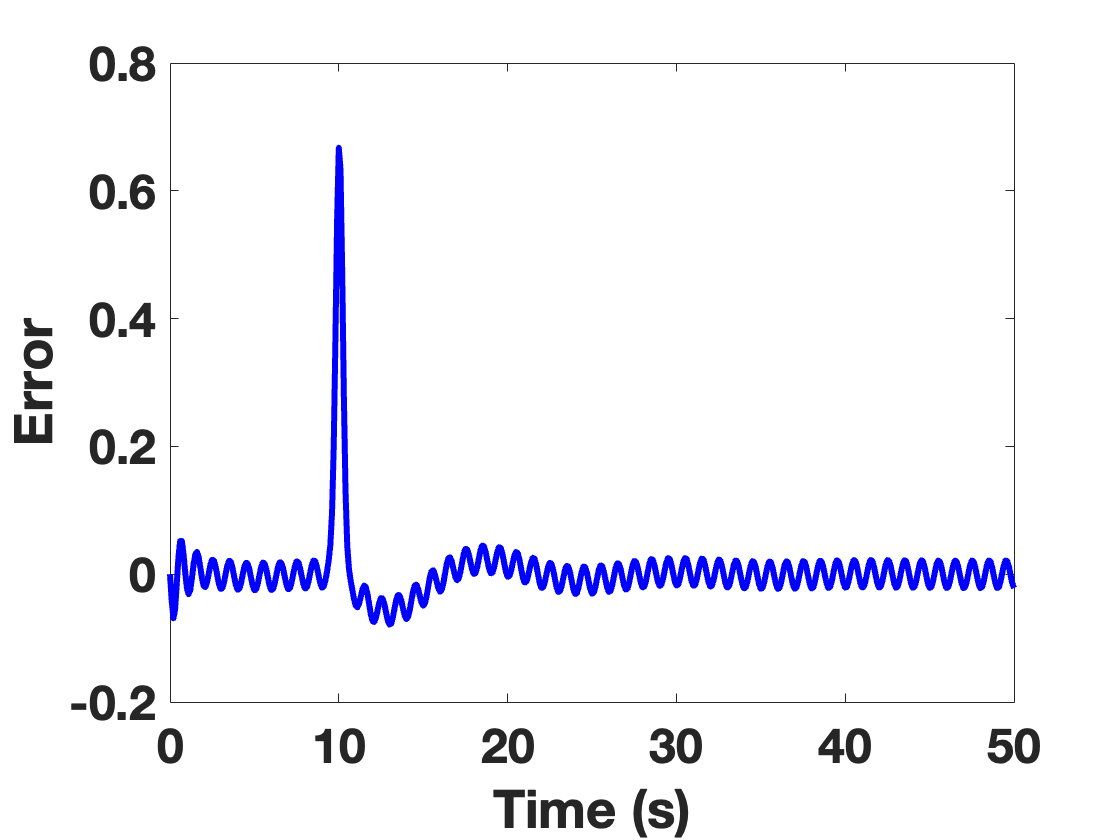}
  \caption{Difference between the outputs of $\Sigma^2_{\text{cart}}$ and $\Sigma^6_{\text{cart}}$.}
  \label{fig:error_cart_gen}
\end{subfigure}
\caption{Comparison of the performance of ROM of a double pendulum controller. The ROM $\Sigma^r$ is obtained by solving $\mathcal{P}_1$.}
\label{fig:gen_cart}
\end{figure}

We denote the model characterized in equation \eqref{eq:dynamics_cart} as $\Sigma^6_{\text{cart}}$. We select $r=2$ and  solve the optimization problem $\mathcal{P}_1$ for the choice of $u_1(t) = 20\sin(2t)$, $u_2(t)=30\sin(2t)$, and $d_1(t)= \frac{30}{0.2\sqrt{2\pi}}\exp\left(-\frac{(t-10)^2}{2(0.2)^2} \right)$ to obtain a reduced order model, denoted by $\Sigma^2_{\text{cart}}$, of $\Sigma^6_{\text{cart}}$. The values of $\gamma^*$ and $\delta^*$ were determined to be $0.0032$ and $0.0042$, respectively. We present the outputs of $\Sigma^6_{\text{cart}}$ and $\Sigma^2_{\text{cart}}$ in Figure \ref{fig:gen_cart}. 

From Figure \ref{fig:gen_cart}, the output of the ROM $\Sigma^2_{\text{cart}}$ obtained by solving the optimization problem $\mathcal{P}_1$ closely matches that of the output of the FOM $\Sigma^6_{\text{cart}}$, even in the presence of disturbance $d_1$ (cf. properties (1) and (3) in Section \ref{subsec:problem}). Further, the state transition matrix $A_2$ of the ROM $\Sigma^2_{\text{cart}}$ was determined to be as follows:
$$A_2=\begin{bmatrix}
    -1.052 & -0.5368\\
    -0.5368 & -0.357
\end{bmatrix}.$$ It can be checked that the matrix $A_2$ obtained as a solution to $\mathcal{P}_1$ is Hurwitz, as guaranteed from Theorem \ref{thm:stability_preserved} (cf. property (2) in Section \ref{subsec:problem}). 


We now show that by combining our approach with moment matching may yield reduced order models with lower approximation error, even when no disturbance acts on the FOM. 
Specifically, we combine the moment matching approach described in \cite{necoara2022optimal} with our approach and solve $\mathcal{P}_{\text{MM}}$ and compare the performance with that in \cite{necoara2022optimal} in which the $\mathcal{H}_2$ norm of the error transfer function is minimized.
Following the approach of \cite{necoara2022optimal}, we fix the observable pair $(S,L)$ to 
\begin{align*}
    S = \begin{bmatrix}
        0 & 1\\0 & 0
    \end{bmatrix}, L = \begin{bmatrix}
        1 & 1
    \end{bmatrix}
\end{align*}
which yields a family of second order models parametrized with $B_2$ that match the first two moments at zero of \eqref{eq:dynamics_cart} with
\begin{align*}
    A_2 = S-B_2L;~ C_2 = \begin{bmatrix}
        1 & -1
    \end{bmatrix}.
\end{align*}

\begin{figure}
\begin{subfigure}{0.48\columnwidth}
  \centering
  \includegraphics[width=\linewidth]{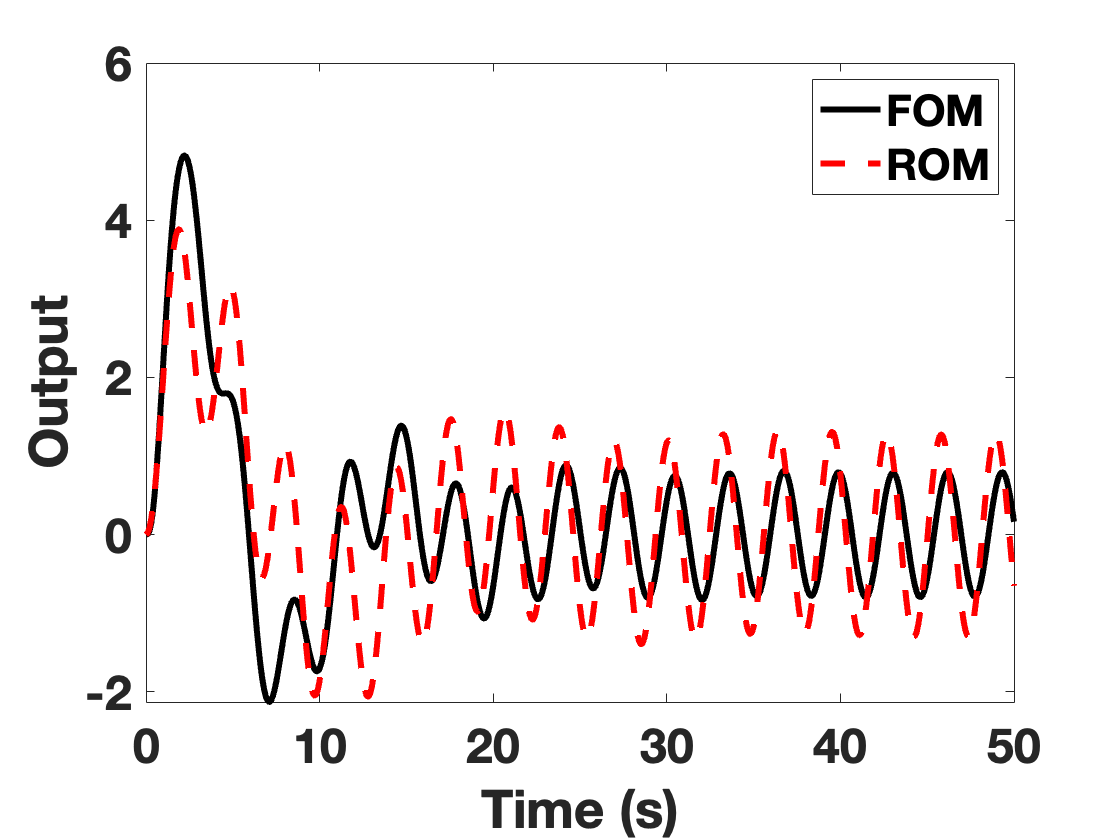}
  \caption{Comparison of outputs of $\Sigma^2_{\text{cart}}$ and $\Sigma^6_{\text{cart}}$. $\Sigma^2_{\text{cart}}$ is obtained as in \cite{necoara2022optimal}.}
  \label{fig:MM}
\end{subfigure}
\hfill
\begin{subfigure}{0.48\columnwidth}
  \centering
  \includegraphics[width=\linewidth]{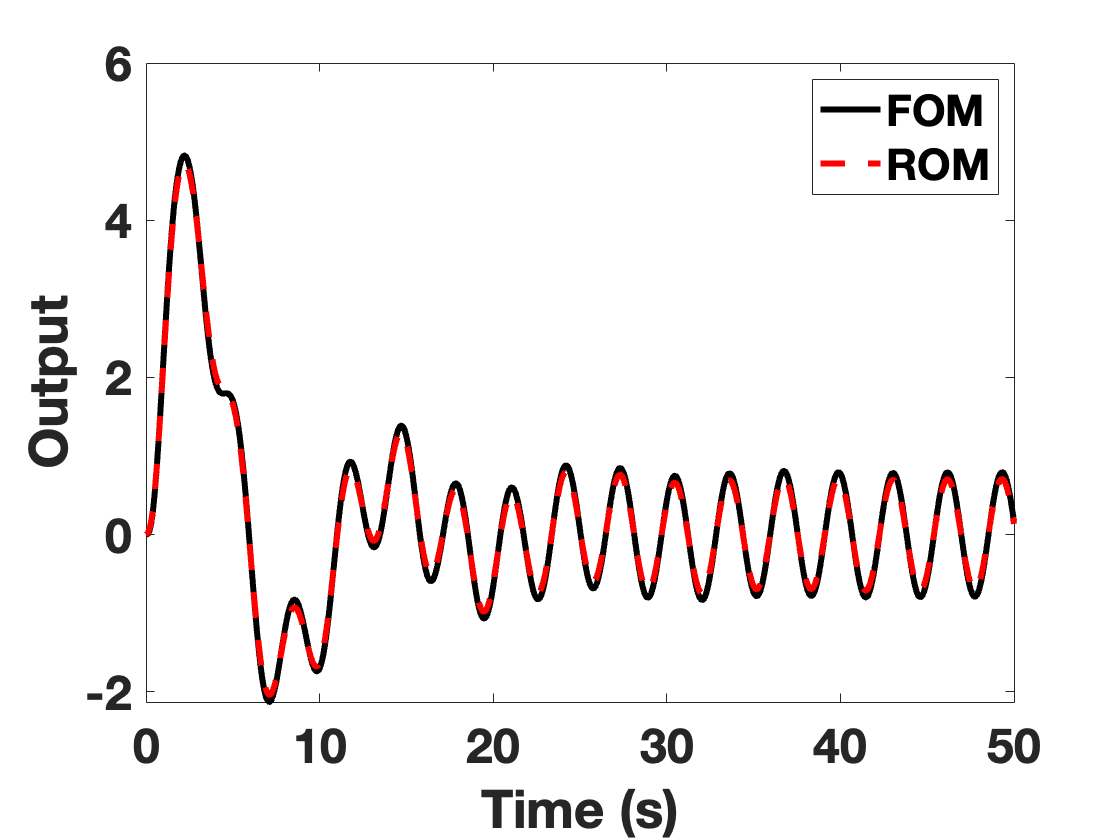}
  \caption{Comparison of outputs of $\Sigma^2_{\text{cart}}$ and $\Sigma^6_{\text{cart}}$. $\Sigma^2_{\text{cart}}$ is obtained from $\mathcal{P}_{\text{MM}}$.}
  \label{fig:error_cart_MM}
\end{subfigure}
\caption{Numerical comparison of the proposed approach with approach based on moment matching \cite{necoara2022optimal} with $d_1(t) = 0$.}
\label{fig:MM_add_on}
\end{figure}


Figure \ref{fig:MM_add_on} presents a comparison of the outputs of the FOM with a ROM obtained by the moment matching approach described in \cite{necoara2022optimal} (with $B_2 = \begin{bmatrix}
    0.2505 & 0.1500
\end{bmatrix}^\top$) and when the ROM is obtained by solving $\mathcal{P}_{\text{MM}}$. 
The inputs $u_1(t)=u_2(t)=20\sin(2t)$ and the disturbance $d_1(t)=0$ are used. The solution obtained by solving $\mathcal{P}_{\text{MM}}$ consists of matrix $B_2=\begin{bmatrix}
    0.6747 & 0.0807
\end{bmatrix}^\top$, $\gamma=0.0052$, and $\delta=0.0028$. Further, the error $\norm{y-y_r}$ was determined to be $0.44$.
Figure \ref{fig:MM_add_on} illustrates that combining our framework with moment matching yields lower approximation errors
as compared to determining a ROM based only on moment matching. This means that lower approximation error may be an additional benefit of our framework apart from the six properties described in Section \ref{subsec:problem}. In particular, combining moment matching with our framework provides an upper bound on the error (cf. \eqref{eq:similarity_ROM}) which may be better as compared to the approach in \cite{necoara2022optimal}.  

\subsection{Spring-Mass-Damper System}
We now consider a coupled spring-mass-damper system with number of masses $N=10$ yielding a system of order $n=20$, and illustrate how our proposed approach can be combined with the balanced truncation (BT) method. Consequently, we will illustrate the robustness as well as the interconnection properties of the reduced order model thus obtained.

Similar to the system in \cite{sato2016structured}, the masses, spring constants, and the viscous friction coefficients were chosen randomly between $1$ and $10$, $1$ and $10^5$, and $0$ and $1$, respectively. We denote the system by $\Sigma^{20}_{\text{SMD}}$. 
We select the order $r=2$ of the ROM and use balanced truncation method to obtain the reduced order model, denoted as $\Sigma^2_{\text{SMD}}$, of $\Sigma^{20}_{\text{SMD}}$. 
Using $\Sigma^2_{\text{SMD}}$ and $\Sigma^{20}_{\text{SMD}}$, we then solve optimization problem $\mathcal{P}_{\text{BT}}$ which yields $\gamma^*=0.122$ and $\delta^*=1.009$. The error $\norm{y-y_r}$ by solving $\mathcal{P}_{\text{BT}}$ was determined to be $0.8076$.

\begin{figure}
    \centering
    \includegraphics[width=0.75\linewidth]{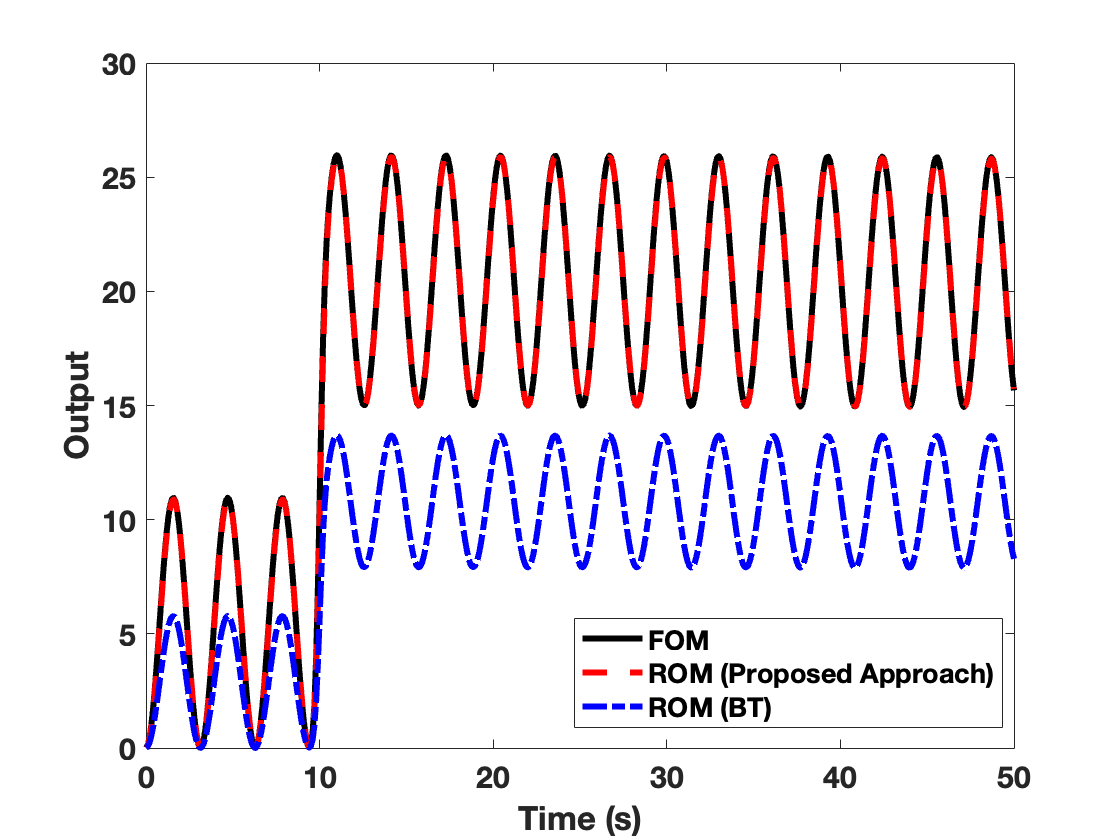}
    \caption{Comparison of output of the spring-mass-damper system and its ROMs when a disturbance acts on the system.}
    \label{fig:SMD_OL}
\end{figure}

Figure \ref{fig:SMD_OL} presents the comparison of the output of the FOM $\Sigma^{20}_{\text{SMD}}$, the ROM obtained by balanced truncation, and the $(\gamma^*,\delta^*)$-ROM obtained by solving $\mathcal{P}_{\text{BT}}$ for the choice of inputs and disturbance $u_1(t) = 11\sin(2t)$, $u_2(t) = 11.1\sin(2t)$, and $d_1(t) = \frac{15}{0.2\sqrt{2\pi}}\exp\left(-\frac{(t-10)^2}{2(0.2)^2} \right)$. 
When balanced truncation is solely used, we first concatenate the system matrices associated with the input and the disturbance and then apply balanced truncation. 
From Figure \ref{fig:SMD_OL}, combining balanced truncation with the proposed approach yields a robust ROM that matches the output of the FOM. 

\begin{figure}
    \centering
    \includegraphics[width=0.75\linewidth]{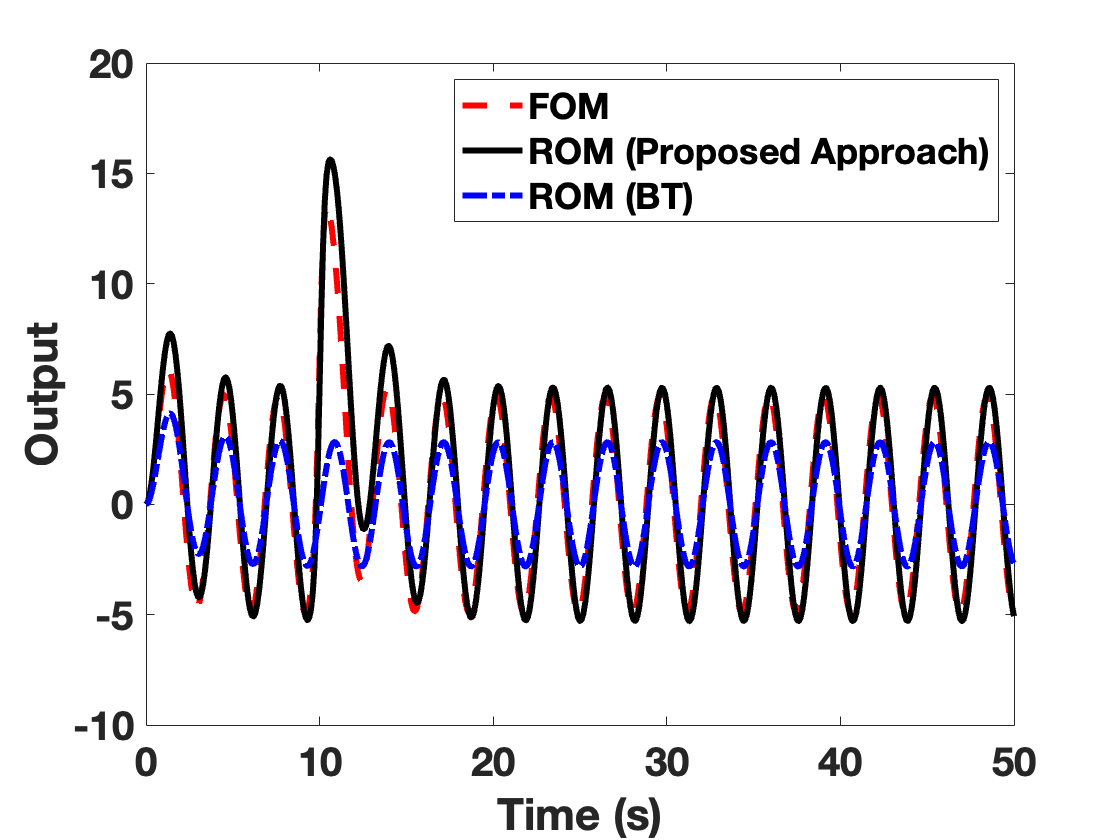}
    \caption{Comparison of output of the spring-mass-damper system and its ROMs when the controller $\mathcal{C}(s)$ is connected in feedback.}
    \label{fig:CL_SMD}
\end{figure}

We now illustrate the closed-loop property of our approach. 
Similarly to \cite{sato2016structured}, we connect the controller $\mathcal{C}(s)=\frac{10^{-5}s+1}{10^{-3}s+1}$ in feedback to the FOM $\Sigma^{20}_{\text{SMD}}$ and the ROM $\Sigma^{2}_{\text{SMD}}$ obtained by solving $\mathcal{P}_{\text{BT}}$.
Figure \ref{fig:CL_SMD} presents the comparison of the output when the controller $\mathcal{C}(s)$ is connected in feedback to (i) the FOM $\Sigma^{20}_{\text{SMD}}$, (ii) the ROM obtained via balanced truncation, and (iii) the $(\gamma^*,\delta^*)$-ROM obtained by solving $\mathcal{P}_{\text{BT}}$.
For the same choice of $u_1(t)$, $u_2(t)$, and $d_1(t)$ as in Figure \ref{fig:SMD_OL}, we observe that the output of the closed-loop system when the controller $\mathcal{C}(s)$ is connected in feedback to the $(\gamma^*,\delta^*)$ ROM closely matches the output of the closed-loop system when the controller $\mathcal{C}(s)$ is connected in feedback to $\Sigma^{20}_{\text{SMD}}$. We further observe that the proposed approach outperforms balanced truncation suggesting the possibility of lower approximation errors by combining our approach with balanced truncation.

\subsection{Building Model}

\begin{figure}
\begin{subfigure}{0.48\columnwidth}
  \centering
  \includegraphics[width=\linewidth]{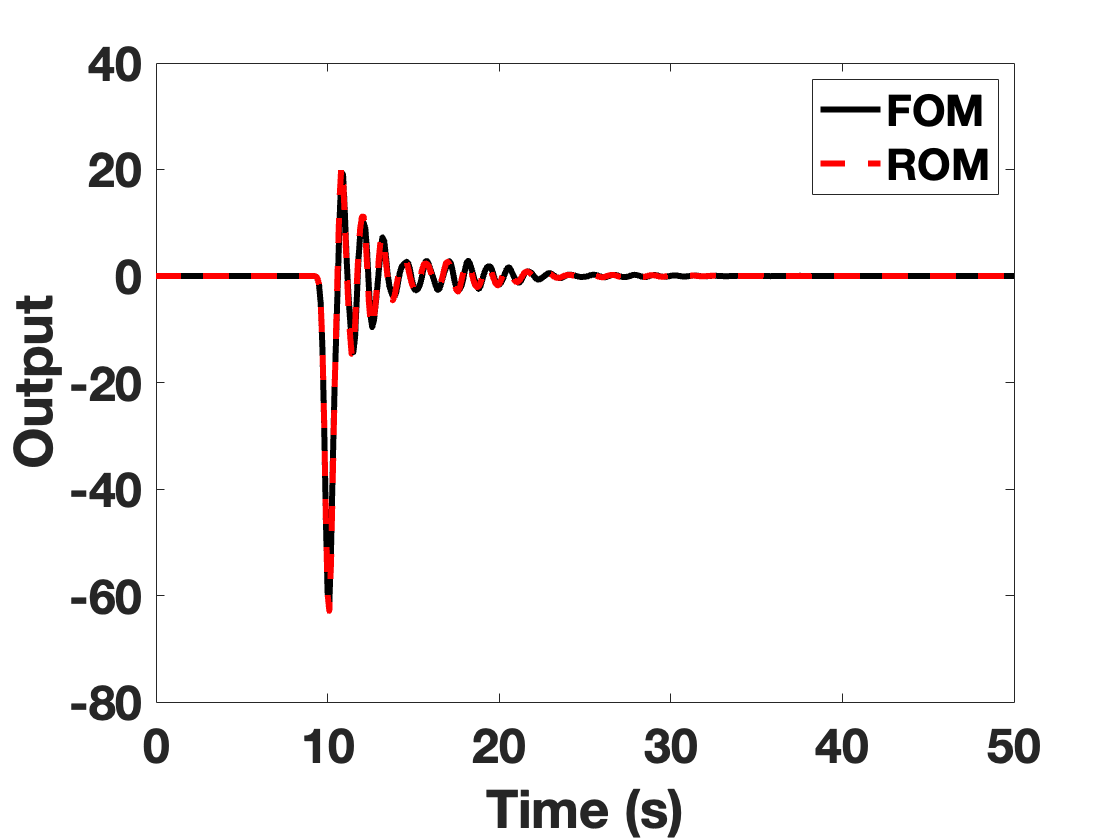}
  \caption{Outputs of $\Sigma^r$ and $\Sigma^n$.}
  \label{fig:output_build}
\end{subfigure}
\hfill
\begin{subfigure}{0.48\columnwidth}
  \centering
  \includegraphics[width=\linewidth]{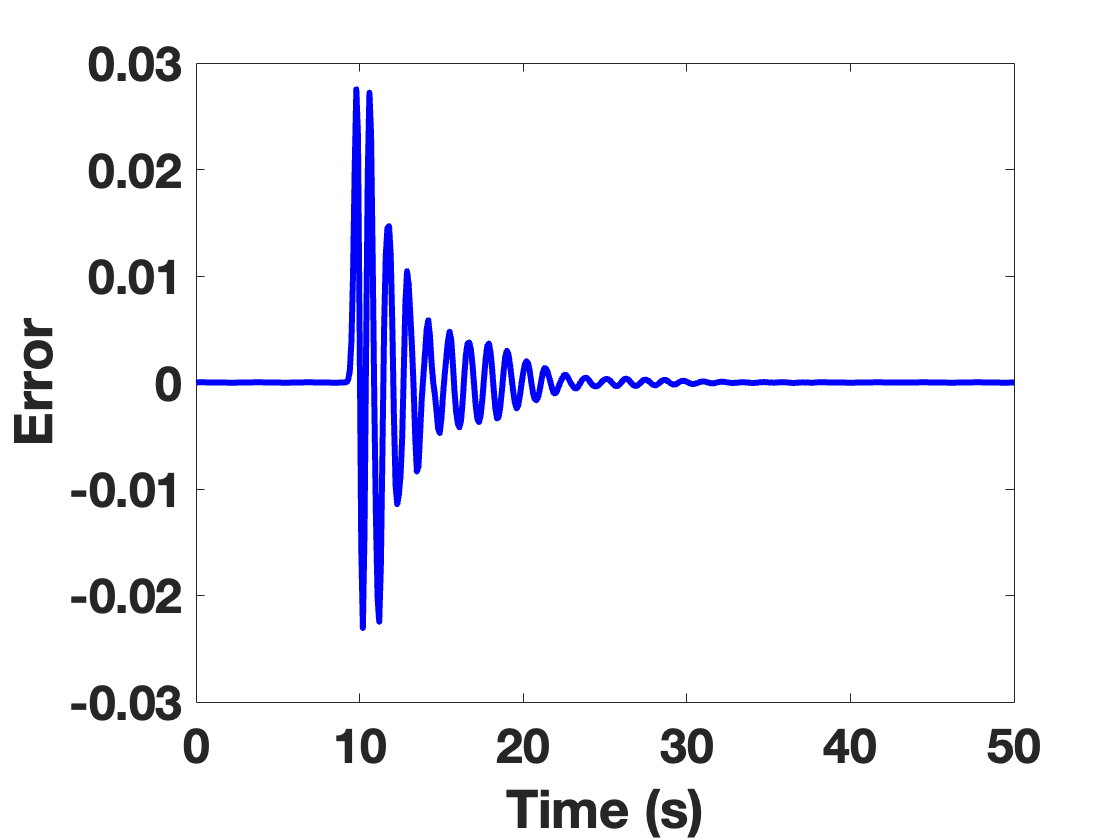}
  \caption{Error between the outputs of $\Sigma^r$ and $\Sigma^n$.}
  \label{fig:error_build}
\end{subfigure}
\caption{Performance of the proposed approach on building model. Disturbance $d_1(t)$ was kept the same as for Figure \ref{fig:error_cart_gen}.}
\label{fig:build}
\end{figure}

Figure \ref{fig:build} presents the proposed framework on a building model\footnote{Online available at http://slicot.org/20-site/126-benchmark-examples-for-model-reduction.}. Specifically, we consider a full order model of the Los Angeles University Hospital. The model is single input and single output, consists of $48$ states, and the pole closest to the imaginary axis has the real part equal to $-2.62 \times 10^{-1}$. We refer to \cite{antoulas2001survey} for more details. By solving optimization problem $\mathcal{P}_{\text{BT}}$ for a given order $r=2$, we obtain a robust ROM of the original building model along with the driving input $d_2$ which ensures that the output of the ROM is similar to that of the FOM. 
In Figure \ref{fig:build}, we see that the output of the ROM obtained by combining our framework with balanced truncation is similar to that of the FOM, even in the presence of an arbitrary disturbance.

\section{Discussion, Conclusion and Future Works}\label{sec:conclusion}
In this work, we provide a framework for model order reduction for continuous-time linear systems such that the obtained ROM continues to be a good approximation even when an arbitrary disturbance acts on the FOM. Additionally, we establish that the proposed framework has the following properties: (1) provable bound on the error, defined as the $\mathcal{L}_2$-norm of the difference between the outputs (2) preserves stability, (3) provable error bound on arbitrary interconnected systems, (4) provable error bound on the output of the FOM when the controller designed for the ROM is applied to it, (5) compatibility with the existing approaches.

We now briefly comment on the applicability of this work. As mentioned in Section \ref{sec:introduction}, the driving input $d_2$ determined using our framework requires the information of the state of the FOM and the ROM. This is because applying $d_2 = Fx$ to the obtained ROM perturbs the dynamics of the ROM such that its output is close to the FOM.
Thus, this work is suitable in reducing high-fidelity digital twins to reduce the computational load. For applications in which the state information from the FOM may not be available, our approach is still applicable by imposing that the driving input $d_2=0$. In this case, the problem reduces to finding an ROM that (without any perturbation) is $(\gamma,\delta)$-similar to the FOM. We believe that the results still hold since substituting $d_2=0$ is a special case.

It must be highlighted that the numerical comparison of our framework with balanced truncation or moment-matching merely suggests that the proposed framework may have comparatively low approximation errors. While we observed consistent results in multiple numerical studies, a definite answer to this comparison remains an open question. Further, while we used specific instances of $u_1$, $u_2$, and $d_1$ for the numerical case studies, we highlight that the error bound (cf. \eqref{eq:similarity_ROM}) holds for every $u_1$, $u_2$, $d_1$. Finally, we note that the optimization problem such as $\mathcal{P}_1$ or $\mathcal{P}_{\text{BT}}$ is only solved once.

Immediate next steps include relaxing the BMI constraint and determining improved error bounds for arbitrary interconnected systems. Future work can also include extending this framework to model reduction of non-linear systems.

\appendices
\section{$(\gamma,\delta)$-Similarity for Unstable Systems}\label{sec:appen_generalization}
In this section, we extend the notion of $(\gamma,\delta)$ similarity to systems with arbitrary initial conditions and which may not be $0$-asymptotically stable.

Given a terminal time $T$, we define the cost function 
\begin{equation}\label{eq:cost_unstable}
    J_T(\tau,x_s,d_2,w) = \int_{\tau}^T |z(t)|^2_R - |w(t)|_Q^2dt,
\end{equation}
where $z(t) = z(t;\tau,x_s,d_2,w)$.
The following result provides an alternate characterization of Definition \ref{def:similarity_def_unstable} in terms of the composite system \eqref{eq:composite_system} and the cost function (\ref{eq:cost_unstable}).
\begin{lemma}
    For $\gamma,\delta>0$, the system $\Sigma_2$ is $(\gamma,\delta)$-similar to system $\Sigma_1$ if and only if there exists matrices $Q, R$, defined in \eqref{eq:Q_and_R_prelim}, and $K\succ 0$ and a constant $\epsilon>0$ such that for all $w\in\mathcal{L}_2$, there exists $d_2\in\mathcal{L}_2$ such that 
    \begin{align}\label{eq:cost_lim_unstable}
        \lim_{T\to\infty} J_T(0,x_s,d_2,w)\leq -\epsilon\norm{w}^2+x_sKx_s.
    \end{align}
\end{lemma}
\begin{proof}
    The proof is analogous to that of \cite[Proposition 2]{pirastehzad2024comparison} and has been omitted for brevity.
\end{proof}
The following result will be instrumental in extending the notion of $(\gamma,\delta)$-similar systems to unstable systems.

\begin{lemma}\label{lem:LQ_stability}
    Suppose that $w(t)=0$, for all $t>0$, yielding the composite system defined in \eqref{eq:composite_system}, as 
    \begin{equation}\label{eq:system_w_0}
    \begin{split}
        \dot{x} &= Ax + Bd_2,\\
        z &= Cx + Dd_2.
    \end{split}
    \end{equation}
    Further suppose that $\Sigma_2$ is $(\gamma,\delta)$-similar to $\Sigma_1$ and that the system \eqref{eq:system_w_0}  does not have any zeros on the imaginary axis.
    Then, there exists a matrix $H\succeq 0$ of the algebraic Riccati equation
    \begin{equation}\label{eq:ARE_LQR_general}
        A^{\top}H+HA-HB(D^{\top}RD)^{-1}B^{\top}H + C^{\top}RC = 0
    \end{equation}
    such that 
    \begin{equation}\label{eq:ARE_LQR_stability_general}
        \text{spec}\left(A-B(D^{\top}RD)^{-1}B^{\top}H \right)\subseteq \mathbb{C}^-.
    \end{equation}
\end{lemma}
\begin{proof}
    From \eqref{eq:cost_lim_unstable} and since $D^{\top}RC = 0$,
    \begin{align*}
        \lim_{T\to\infty} J_T(0,x_s,d_2,w) &= \norm{z}_R^2 \\
        &= \int_0^{\infty}x^{\top}C^{\top}RCx + d_2^{\top}D^{\top}RDd_2 \mathrm{d}t.
    \end{align*}
    As $\Sigma_2$ is $(\gamma,\delta)$-similar to $\Sigma_1$, it follows from \eqref{eq:cost_lim_unstable} that there exists a $d_2\in\mathcal{L}_2$ such that $\lim_{T\to\infty} J_T(0,x_s,d_2,w)$ is bounded.
    Following analogous steps as in \cite[Theorem 10.13]{trentelman2002control}, it follows that there exists a real positive definite solution $H$ of \eqref{eq:ARE_LQR_general}. 

    We will now establish that the solution of \eqref{eq:ARE_LQR_general} satisfies \eqref{eq:ARE_LQR_stability_general}. Let $F\coloneqq -\left(D^{\top}RD \right)^{-1}B^{\top}L$ Since $D^{\top}RC=0$, \eqref{eq:ARE_LQR_general} can be rewritten as
    \begin{align}\label{eq:ARE_LQR_modified}
        (A+BF)^{\top}H &+ H(A+BF)  \nonumber\\
        &+ (C+DF)^{\top}R(C+DF) = 0.
    \end{align}
    Let $v$ (resp. $\lambda$) denote the eigenvector (resp. eigenvalue) of $A+BF$. Then multiplying \eqref{eq:ARE_LQR_modified} from the left and right with $v^*$ and $v$, respectively, yields $2\text{Re}\{\lambda\}v^*Hv = -|(C+DF)v|_R^2\leq 0.$
    It follows that $\text{Re}\{\lambda\}\leq 0$ unless $Hv=0$. $Hv=0$ or $\text{Re}\{\lambda\}=0$ implies that $(C+DF)v=0$ which further implies that $\lambda$ is a zero of the system (see Exercise 10.1 and Exercise 13.3 in \cite{trentelman2002control}). Given the assumption that the zeros of the system do not lie on the imaginary axis, it follows that $\text{Re}\{\lambda\} < 0$, i.e., $\text{spec}\left(A-B(D^{\top}RD)^{-1}B^{\top}L \right)\subseteq \mathbb{C}^-.$
\end{proof}

\begin{theorem}\label{thm:similarity_unstable_riccati}
    Suppose that the composite system \eqref{eq:composite_system} does not have any zeros on the imaginary axis. For $\gamma,\delta>0$, system $\Sigma_2$ is $(\gamma,\delta)$-similar to $\Sigma_1$ if and only if there exist positive constants $\eta,\mu$ and a positive definite matrix $K$ such that the following algebraic Riccati equation
    \begin{equation}\label{eq:ARE}
    \begin{split}
        0 = A^\top P+PA+C^\top RC+PEQ^{-1}E^\top P-\\
        PB(D^\top RD)^{-1}B\top P
    \end{split}
    \end{equation}
    admits a positive semi-definite solution $P$ such that 
    \begin{equation}\label{eq:stability_unstable}
        \text{spec}\left(A-B(D^\top RD)^{-1}B^{\top} P+EQ^{-1}E^{\top} P\right)\subseteq \mathbb{C}^-.
    \end{equation}
\end{theorem}
\begin{proof}
    We will first establish the \emph{only if} part, i.e., if $\Sigma_2$ is $(\gamma,\delta)$-similar to $\Sigma_1$, then a solution  $P\succeq 0$ of \eqref{eq:ARE} exists which satisfies \eqref{eq:stability_unstable}.
    
    \textit{Only if:} 
    Recall from Lemma \ref{lem:LQ_stability} that there exists a solution $H$ of \eqref{eq:ARE_LQR_general}.
    Define the cost function as follows which utilizes the matrix $L$ as an end point penalty.
    \begin{align*}
        \mathcal{J}_T(0,x_s) 
        & = \sup_{w \in \mathcal{L}_2}\inf_{d_2\in\mathcal{L}_2} \left(\norm{z}_{[0,T],R}^2 - \norm{w}^2_{[0,T],Q} \right. \\
        & \left.  + x^{\top}(T) H x(T) \right)\\
        & \leq \sup_{w \in \mathcal{L}_2}\inf_{d_2\in\mathcal{L}_2} \left(\norm{z}_{[0,T],R}^2 - \norm{w}^2_{[0,T],Q} + \right .\\
        &\left . \{ \inf_{d_2\in\mathcal{L}_2[T,\infty]}\int_T^\infty|z(t)|_R^2\mathrm{d}t \mid \forall t> T: w(t)=0, \right.\\
        & \left. \lim_{t\to\infty}x(t) = 0 \}\right).
    \end{align*}
    Since the term in the second line of the last inequality is bounded as a result of linear quadratic optimal control, choosing $d_2$ according to \eqref{eq:cost_lim_unstable} yields
    \begin{align*}
        &\mathcal{J}_T(0,x_s) \leq \sup_{w\in \mathcal{L}_2}\inf_{d_2\in\mathcal{L}_2} \{\norm{z}_R^2-\norm{w}_Q^2 \mid \forall t>T:w(t) = 0\}\\
        &\leq \sup_{w\in\mathcal{L}_2}\{\norm{z}^2_R-\norm{w}_Q^2\}\leq \sup_{w\in\mathcal{L}_2}\{x_s^\top Kx_s -\epsilon\norm{w}^2\},
    \end{align*}
    where the last inequality is obtained from equation \eqref{eq:cost_unstable}.

    This implies that $\mathcal{J}_T$ is bounded from above uniformly with $T$. Now consider the following differential Riccati equation with $P(0)=L$:
    \begin{equation}\label{eq:DRE}
    \begin{split}
        \dot{P} =& A^{\top}P + PA + C^{\top}RC+PEQ^{-1}E^{\top}P \\
        &- PB(D^{\top}RD)^{-1}B^{\top}P; \quad P(T) = 0.
    \end{split}
    \end{equation}
    From \cite[Theorem 10.7]{trentelman2002control}, there exists a $T_1\geq 0$ such that \eqref{eq:DRE} has a solution $P$ on the interval $[0,T_1]$ with $P(T)=0$. Following analogous steps to that in \cite[Lemma 13.5]{trentelman2002control}, we use completion of the squares argument and obtain that 
    \begin{align*}
        \mathcal{J}_T(\tau,x_s) = x_s^{\top}P(\tau)x_s, \quad \forall \tau\in[0,T_1].
    \end{align*}
    This implies that $P$ is uniformly bounded. Now, using analogous arguments as in \cite[Theorem 13.3]{trentelman2002control}, we conclude that a solution $P$ with $P(0)=H$ exists on the complete interval $[0,\infty)$, $P(t)\to \bar{P}$ as $t\to\infty$, $\bar{P}\geq H$, and that $\bar{P}$ satisfies the algebraic Riccati equation \eqref{eq:ARE} for $P=\bar{P}$. 

    Let $F\coloneqq -(D^{\top}RD)^{-1}B^{\top}P$. Then, following analogous steps as in \cite[Theorem 13.3]{trentelman2002control}, we establish that $A+BF$ is Hurwitz. 
    Now define
    \begin{align*}
        \bar{w}_T(t) \coloneqq \begin{cases}
            w_T(t), &t\leq T,\\
            0, &t>T,
        \end{cases}
    \end{align*}
    where $w_T(t)\coloneqq Q^{-1}E^{\top}Px(t)$. Follow similar steps as in \cite[Theorem 1]{pirastehzad2024comparison} and using \eqref{eq:cost_lim_unstable} yields
    \begin{align*}
        J_T(0,x_s,Fx,w_T) &\leq \norm{z(\cdot;0,x_s,Fx,\bar{w}_T)}_R^2 - \norm{\bar{w}_T}^2_Q\\
        & \leq -\epsilon\norm{\bar{w}_T}^2+x_sKx_s.
    \end{align*}
    From this point on, the proof is analogous to that in \cite[Theorem 1]{pirastehzad2024comparison} and has been omitted for brevity.

    \textit{If part:} Select $d_2^* = Fx$, where $F = -(D^{\top}RD)^{-1}B^{\top}P$. We begin by establishing that $(A+BF)$ is Hurwitz. The idea is to prove through contradiction.

    Suppose that $A+BF$ is not Hurwitz and let $v$ (resp. $\lambda$) denote an eigenvector (resp. eigenvalue) of $A+BF$. Since $D^{\top}RC = 0$, \eqref{eq:ARE} can be written as
    \begin{equation}\label{eq:modified_ARE}
    \begin{split}
        0 &= (A+BF)^{\top}P + P(A+BF) \\
        &+ (C+DF)^{\top}R(C+DF) + PEQ^{-1}E^{\top}P.
    \end{split}
    \end{equation}
    Then, multiplying \eqref{eq:modified_ARE} by $v^*$ and $v$ from the left and right, respectively, yields
    \begin{align}\label{eq:eigen}
        2\text{Re}\{\lambda\}v^*Pv = -|E^{\top}Pv|_{Q^{-1}}^2-|(C+DF)v|^2_R\leq 0.
    \end{align}
    Given the assumption that $\text{Re}\{\lambda\} \geq 0$, \eqref{eq:eigen} holds either if $Pv=0$ or $\text{Re}\{\lambda\} = 0$.
    For either case, since $|E^{\top}Pv|_{Q^{-1}}^2$ and $|(C+DF)v|^2_R$ are non-negative, it follows that $E^{\top}Pv = 0$ which is not possible since \eqref{eq:stability_unstable} holds. This implies that $A+BF$ must be Hurwitz which further implies, from \cite[Lemma 4.8]{lozano2013dissipative}, that $d_2^*\in \mathcal{L}_2$. Next, using completion of squares argument and selecting $K=P(0)+\tilde{\epsilon}\mathbf{I}$, it follows that 
    \begin{align*}
        \lim_{T\to\infty} J_T(0,x_s,d_2^*,w) &\leq x_sKx_s \\
        &- \int_0^{\infty}|w(t)-Q^{-1}E^{\top}P(t)x(t)|_Q^2\mathrm{d}t.
    \end{align*}
    From this point on, the proof is analogous to that in \cite[Theorem 1]{pirastehzad2024comparison}.
\end{proof}

Theorem \ref{thm:similarity_unstable_riccati} provides an algebraic characterization of $(\gamma,\delta)$-similarity for unstable systems with arbitrary initial conditions. However, to determine the constants $\eta$ and $\mu$ from \eqref{eq:ARE} can be challenging. Recall that the notion of $(\gamma,\delta)$-similarity merely requires the existence of these constants. Thus, utilizing the techniques from dissipative theory and analogous to \cite{pirastehzad2024comparison}, we now aim to characterize a simple verifiable condition for the notion of $(\gamma,\delta)$-similarity. We first establish that $d_2$ can also be obtained through state feedback.

\begin{lemma}\label{lem:diss_interm_lem}
    For $\gamma,\delta>0$, $\Sigma_2$ is $(\gamma,\delta)$-similar to $\Sigma_1$ if and only if there exist constants $\epsilon, \eta, \mu>0$ and matrices $F$ and $K\succ 0$ such that the composite system
    \begin{equation}\label{eq:composite_diss_system}
        \begin{split}
            \dot{x} &= \left(A + BF\right)x + Ew,\quad x(0) = x_s\\
            &z = \left(C + DF\right)x
        \end{split}
    \end{equation}
    is $0$-asymptotically stable and satisfies
    \begin{align}\label{eq:cost_unstable_dissipative}
        \forall w\in \mathcal{L}_2: \norm{z}_R^2 - \norm{w}_Q^2\leq -\epsilon\norm{w}^2 + x_sKx_s.
    \end{align}
\end{lemma}
\begin{proof}
    The proof is analogous to that of \cite[Lemma 3]{pirastehzad2024comparison} and has been omitted for brevity.
\end{proof}

\begin{lemma}
    For $\gamma,\delta>0$, $\Sigma_2$ is $(\gamma,\delta)$-similar to $\Sigma_1$ if there exist constants $\epsilon, \eta, \mu>0$ and matrices $F$ and $K\succ 0$ such that the composite system \eqref{eq:composite_diss_system} is $0$-asymptotically stable and strictly dissipative with respect to the supply rate
    \begin{equation}\label{eq:supply_rate}
        s(w,z) = \begin{bmatrix}
            w\\
            z
        \end{bmatrix}^{\top}\begin{bmatrix}
            Q & \mathbf{0}\\
            \mathbf{0} & -R
        \end{bmatrix}\begin{bmatrix}
            w\\ z
        \end{bmatrix},
    \end{equation}
    i.e., there exists a function $V:\mathbb{R}^n\to [0,\infty)$ and an $\epsilon>0$ such that 
    \begin{equation}\label{eq:diss_func}
        V(x(t_1))\leq V(x(t_0)) + \int_{t_0}^{t_1} s(w(t),z(t)) \mathrm{d}t - \epsilon\int_{t_0}^{t_1}|w(t)|^2\mathrm{d}t,
    \end{equation}
    for all $t_0\leq t_1$ and all signals $x, w,$ and $z$ that satisfy \eqref{eq:composite_diss_system}.
\end{lemma}
\begin{proof}
    Suppose that constants $\epsilon, \eta, \mu>0$ and matrices $F$ and $K\succ 0$ exist such that $A+BF$ is $0$-asymptotically stable. Then, for $t_0=0$, $x(0)=x_s$, \eqref{eq:diss_func} can be rewritten as
    \begin{align*}
        \norm{z}_R^2-\norm{w}_Q^2\leq V(x_s) - V(x(t_1)) - \epsilon\norm{w}^2.
    \end{align*}
    Selecting $V = x^{\top}(t)Kx(t)$ yields
    \begin{align*}
        \norm{z}_R^2-\norm{w}_Q^2 &\leq x_s^{\top}Kx_s - x(t_1)^{\top}Kx(t_1) - \epsilon\norm{w}^2\\
        &\leq x_s^{\top}Kx_s - \epsilon\norm{w}^2
    \end{align*}
    which establishes the claim.
\end{proof}

\section{Additional Details for Motivating Example}\label{app:motivation_details}
In this section, we provide details such as the system matrices, the input signal, and the disturbance used to construct the motivating example in \ref{fig:not_robust} in Section \ref{sec:introduction}.

For Figure \ref{fig:BT_not_robust}, we use the following system matrices:
\begin{align*}
    &A = \begin{bmatrix}
        0 & 1 & 0 & 0\\
        -2 & -0.5 & 0.1 & 0\\
        0.1 & 0 & 0 & 1\\
        0 & 0 & -2 & -0.1
    \end{bmatrix}, 
    B =\begin{bmatrix}
        1 & 0 & 0 & 0
    \end{bmatrix}^\top,\\
    & C = B^\top.
\end{align*}
The input $u(t)$ and disturbance $d(t)$ were selected as $20\sin(2t)$ $d(t) = \tfrac{30}{0.2\sqrt{2\pi}}\exp{\frac{-(t-10)^2}{2*0.2^2}}$, respectively. 

For Figure \ref{fig:MM_not_robust}, we consider the double-pendulum model and refer the reader to Section \ref{sec:numerics} for details on the system matrices. We used the ROM obtained in \cite{necoara2022optimal} for our example. Specifically, the reduced order model obtained is
\begin{align*}
    &A_r = \begin{bmatrix}
        -0.3354 & 0.6486\\
        -0.3250 & -0.3060
    \end{bmatrix},
    B_r = \begin{bmatrix}
        0.3467 & 0.3160
    \end{bmatrix}^\top\\
    & C_r = \begin{bmatrix}
        1.0049 & -1.0832
    \end{bmatrix}
\end{align*}
Further, the choice of input $u(t)$ and $d(t)$ was chosen to be the same as for Figure \ref{fig:BT_not_robust}.


\bibliographystyle{ieeetr}
\bibliography{references}
\end{document}